\documentclass[letterpaper,12pt,leqno]{article}
\usepackage{paper}

\hypersetup{pdftitle={Paper Example}}

\usepackage{paralist}

\newcommand{\bv}{{\bf V}}
\newcommand{\bu}{{\bf U}}

\newcommand{\ubw}{\underline{{\bf W}}}

\newcommand{\tty}{{\mathbf{Y}_0}}
\newcommand{\bye}{\mathbf{y}_1}

\newcommand{\byt}{\mathbf{y}_2}
\newcommand{\bytt}{\mathbf{y}_2^\top}

\newcommand{\ubww}{\underline{\mathbf{w}}}

\newcommand{\stag}{{\rm stag}}
\newcommand{\block}{{\rm block}}
\newcommand{\cluster}{{\rm cluster}}
\newcommand{\single}{{\rm single}}
\newcommand{\gen}{{\rm gen}}
\newcommand{\fat}{{\rm fat}}
\newcommand{\tall}{{\rm thin}}
\newcommand{\squ}{{\rm square}}
\newcommand{\last}{{\rm last}}
\newcommand{\twfe}{\mathrm{TWFE}}
\newcommand{\treat}{{\rm tr}}
\newcommand{\control}{{\rm co}}
\newcommand{\post}{{\rm post}}
\newcommand{\pre}{{\rm pre}}
\newcommand{\ttock}{1}
\newcommand{\ttick}{0}
\newcommand{\htautwfe}{\hat\tau^{\twfe}}

\newcommand{\by}{{\bf Y}}
\newcommand{\bw}{{\bf W}}

\newcommand{\indep}{\perp\!\!\!\perp}
\newcommand{\pr}{{\rm pr}}

\newcommand{\oy}{\overline{Y}}
\newcommand{\did}{\mathrm{DID}}
\newcommand{\mme}{{\mathbb{E}}}

\begin{document}

\title{Causal Models for Longitudinal and Panel Data: A Survey}
\author{Dmitry Arkhangelsky \& Guido Imbens
\thanks{Dmitry Arkhangelsky, darkhangel@cemfi.es,  CEMFI; Guido Imbens, imbens@stanford.edu, Stanford University.
This paper originated in the Sargan lecture presented online at the 2021 Royal Economic Society Meetings. We are grateful for the comments by Manuel Arellano, Apoorva Lal, Ganghua Mei, the editor, Jaap Abbring and two reviewers. We thank the Office of Naval Research for support under grant numbers N00014-17-1-2131 and N00014-19-1-2468 and Amazon for a gift.}}
\date{June 2024}                       
\begin{titlepage}
\maketitle
 In this survey we discuss the recent causal panel data literature. This recent literature has focused on credibly estimating causal effects of binary interventions in settings with longitudinal data, emphasizing practical advice for empirical researchers. It pays particular attention to heterogeneity in the causal effects, often in situations where few units are treated and with particular structures on the assignment pattern. The literature has extended earlier work on difference-in-differences or two-way-fixed-effect estimators. It has more generally incorporated factor models or interactive fixed effects. It has also developed novel methods using synthetic control approaches.

\vskip0.7cm

\end{titlepage}

\section{Introduction}

In recent years, there has been a fast-growing and exciting body of research on new methods for estimating causal effects in panel or longitudinal data settings where we observe outcomes for a number of units repeatedly over time. This literature has taken some of the elements of the earlier panel data literature and combined them with insights from the causal inference literature. It has largely focused on the case with binary treatments, although the insights obtained in this body of work extend beyond that setting. Much of this work focuses on settings where traditionally Difference-In-Differences (DID) and Two-Way-Fixed-Effect (TWFE) methods (we largely use the two terms interchangeably, primarily using the TWFE acronym) have been popular among empirical researchers. In this survey, we review some of the methodological research and make connections to various other parts of the panel data literature. 

Although we intend to make this survey of interest to empirical researchers, it is not primarily a guide with recommendations for specific cases. Rather, we intend to lay out our views on this literature in order that practitioners can decide which of the methods they wish to use in particular settings. In line with most of the literature, we see the models and assumptions used in this literature not as either holding exactly or not holding, but as approximations that may be useful in particular settings. For example, the TWFE setup has been criticized as making assumptions that are too strong. At some level, that is true almost by definition: parallel trends are unlikely to hold over any extended period of time for a large number of units.  Similarly, assuming the absence of dynamic effects is unlikely to ever hold exactly, and treatment effects are surely heterogeneous. Nevertheless, in many cases, fixed-effect models with time-invariant constant treatment effects may be effective baseline models. Understanding when those models are adequate, when relaxing them is likely to improve estimation,  and what useful generalizations to relax their underlying assumptions are, is what we intend to do in this survey. 

Ultimately, and this is perhaps our strongest statement on the relative merits of the various methods, we recommend against the current routine use of the standard TWFE estimator or related estimators. 
These methods have been very popular in empirical work and, in fact, continue to increase in popularity. See \citep{goldsmith} for a discussion tracing trends in their usage in recent years.
Nevertheless, there are now many methods that generalize this estimator that we view as more attractive in practice. Some of these are based on strictly more general models for the potential outcomes (in particular factor models). Other use local versions of the TWFE estimator through weights on the cross-sectional and/or time dimension (building on the synthetic control literature). Both approaches often use regularization to ensure good performance by avoiding overfitting  even when the simpler methods ({\it e.g.,} based on the standard TWFE model) are adequate. Although both approaches, more general outcome models and weights, are in our view superior to the standard TWFE methods, their relative performance varies by context and the relative merits are the subject of ongoing research. We therefore offer no specific recommendations for any one particular method. In addition, the more general methods still share some of the unattractive features of the TWFE estimator.  In particular they often pay little explicit attention to dynamics and time-series structure in potential outcomes.
In addition many of the methods pay limited attention to the assignment mechanism.
Developing methods that address these concerns is a promising and practically relevant area of future research.

A second recommendation concerns 
some of the specific issues raised in the recent TWFE literature. This literature has generated valuable new insights into the complications raised by the presence of heterogeneity in treatment effects. These insights have improved our understanding of the challenges with panel data. The new estimators developed in this literature do not, however, in our view, fully address all the practical challenges. On the positive side, they allow for much more heterogeneity in treatment effects than the earlier panel literature.  On the negative side, like the earlier TWFE literature, the proposed estimators rely on unrealistically strong additivity and linearity assumptions on the potential outcomes, limiting the credibility of the estimates of counterfactuals. In our view there needs to be more balance in the richness of the models for the control potential outcomes and the richness of the models for the treatment effects. Putting no structure on the heterogeneity of the treatment effects is at odds with the typical goal of predicting the effect of implementations of the new policies to other locations, populations, or time periods rather than simply evaluating those policies on the currently exposed populations.

The paper is organized as follows.
After the introduction, we first discuss in Section \ref{section:traditional} some of the earlier econometric panel data literature. This serves both to set the stage for the framing of the questions of the current literature as well as to clarify differences in emphasis between the traditional and new literatures. We also point out that some important conceptual issues that had been raised in the earlier literature have received less attention recently and that some are even in danger of being entirely ignored in the current literature.

Next, in Section \ref{section:configurations}, we discuss three ways of organizing the panel data literature. First we consider a classification  by types of data available, {\it e.g.,} proper panel data, repeated cross-sections, or row and column exchangeable data.  (The latter  refers to a matrix of data where both rows and columns are exchangeable, similar to panel data without any time-series structure.) Second we discuss an organization by shapes of the data frame, {\it e.g.,} many units or many periods. Finally we discuss  a classification based on the assignment for the causal variable of interest, {\it e.g.,} block assignment, single treated unit, single treated period, or staggered adoption. We find these  classifications useful because they matter for the relevance of various methods that have been proposed and they help organize them. Although the earlier econometric panel data literature also stressed the importance of the relative magnitude of the time and unit dimension as we do in our second classification, the realization that the structure of the assignment process is important is a more recent insight.
Many of the recent papers focus on particular parts of the general space of panel data inferential problems. For example, the vast literature assuming unconfoundedness in panel data settings has focused largely on the setting with a large number of units and relatively few time periods, and a subset of the units treated in the last period. In contrast, the Synthetic Control (SC) literature has primarily focused on the setting where the cross-section and time series dimension are comparable in size, and where one or few units are treated from some period onwards. The recent DID/TWFE literature has paid particular attention to the setting with staggered adoption patterns in the assignment. The singular focus of some of these literatures has helped in advancing them more rapidly, but occasionally, insights from related settings have been overlooked.

In Section \ref{notation} we introduce some of the notation and estimands. We use, as in much of the causal inference literature, the potential outcome notation that makes explicit the causal nature of the questions.

In Section \ref{section:twe} we introduce the standard DID/TWFE setup as a stepping stone to the discussion of the recent developments in causal panel data literature. We see four main threads in the new causal panel literature, which we discuss in Sections \ref{section:staggered} through \ref{section:future}.

First,  in Section \ref{section:staggered}, we discuss the staggered adoption case. 
Much of the earlier TWFE literature concentrated on the case with a common adoption date, 
In contrast, one strand of the recent literature has focused on the setup where different groups adopt treatments at different points in time. In this staggered adoption case, recent research has highlighted some specific concerns with the standard TWFE estimator. In particular, in cases with general treatment effect heterogeneity, the implicit negative weights on the building blocks of the TWFE estimator have been argued to be unattractive, and alternatives have been proposed. We argue that these concerns have perhaps been exaggerated.

Second, as discussed in  Section \ref{section:relax}, the recent literature has generalized the popular TWFE structure to factor models. An important part of this literature is the SC approach developed in a series of influential papers by Alberto Abadie and coauthors (\citealp*{abadie2003, abadie2010synthetic}). Although this literature shares key features with the TWFE literature, it has largely developed separately, ignoring some of the gains that can arise from combining the insights from each of them. 

In the third strand, we consider in Section \ref{section:nonlinear} a different class of generalizations of the TWFE setup,  allowing for nonlinear models.

 Fourth, as discussed in Section \ref{section:design}, the modern causal panel literature has sometimes taken a design-based approach to inference where the focus is on uncertainty arising from the assignment mechanism rather than a model-based or sampling-based perspective that is common in the earlier literature. 

In Section \ref{section:future}, we discuss open questions in the causal panel data literature which we view as exciting avenues for future research.

Finally, in Section \ref{recommendations}, we discuss some recommendations for empirical practice.
 
 There are some excellent recent discussions of the new DID/TWFE and causal panel data literature that are complementary to this survey. They differ in their focus and in the perspectives of the authors and complement ours in various ways.
  Some of these surveys (\citealp{de2023two, roth2023s}) focus more narrowly on the DID/TWFE setting with heterogeneous treatment effects. They do not stress the connections with the synthetic control methods and factor models that we view as an important feature of the current panel data literature. In contrast, \citep{abadie2019using} focuses primarily on synthetic control methods. In the current survey, we stress deeper linkages between these ideas and the TWFE literature  as well as the potential benefits of combining them. Recent surveys in the political science literature, \citep{liu2022practical, xu2023causal}, more in line with the current survey, also discuss the connections between synthetic control, unconfoundedness, and TWFE approaches.

In this discussion, we use a number of acronyms. For reference, we list those that we use regularly in Table 1.
\begin{table}\label{tabel1}
\begin{centering}\textsc{Table 1: Acronyms}
\par\end{centering}
\centering{}\label{tabel_classification}\vskip0.3cm \centering{}%
\begin{tabular}{ll}
 \tabularnewline
\hline 
DID & Difference In Differences\\
TWFE & Two Way Fixed Effect\\
SC & Synthetic Control\\
GRCS & Grouped Repeated Cross Section\\
RCED & Row Column Exchangeable  Data \\
SDID & Synthetic Difference in Differences\\
CIC & Changes in Changes\\
NNMC & Nuclear Norm Matrix Completion
\tabularnewline
\end{tabular}
\end{table}

\section{The Econometrics Panel Data Literature}\label{section:traditional}

Although the new panel literature ostensibly focuses on different estimands and settings and emphasizes different concerns about internal and external validity, many of the methods are closely related to those discussed in the earlier econometric panel data literature. Here we discuss at a high level some of the key insights from the earlier literature, in so far they relate to the current literature,  and some marked differences between the two. We come back to some of the specific areas of overlap in later sections. We do not attempt to review the earlier econometric literature, partly because that is a vast literature in itself, but mainly because there are many excellent surveys and textbooks, including \citep{arellano2001panel, arellano2003panel,baltagi2008econometric,hsiao2022analysis, wooldridge2010econometric,arellano2011nonlinear}.

First of all, by the econometric panel data literature, we mean primarily the literature 
from the 1980s to the early 2000s, as for example, reviewed in the surveys and textbooks, including
\citep{chamberlain1982multivariate, chamberlain1984panel, hsiao2022analysis, arellano2003panel, arellano2001panel, baltagi2008econometric, wooldridge2010econometric, arellano2011nonlinear}. This literature was initially motivated by the increased availability of various large public-use longitudinal data sets starting in the 1960s. These data sets included the Panel Study of Income Dynamics, the National Longitudinal Survey of Youth, which are proper panels where individuals are followed over fairly long periods of time, and the Current Population Survey, which, although primarily a repeated cross-section data set, has some short-term panel features, and at the state level can be viewed as a longer panel data set. 
These data sets vary substantially in the length of the time series component, motivating different methods that could account for such data configurations.

The primary focus of the econometric literature has been on estimating invariant or structural parameters in the sense of \citep{goldberger1991course}. Part of the literature analyzed fully parametric models, but more often semiparametric settings were considered. The parameters of interest could be causal in the modern sense, but the term itself would rarely be used explicitly. A major concern in this literature has been the presence of time-invariant unit-specific components. The literature distinguished between two types of such components: first, the so-called fixed effects and random effects. Fixed effects were conditioned on in the analyses and were modeled as unrestricted in their correlation with other variables. Random 
effects  were treated as stochastic and often assumed to be uncorrelated with observed covariates (though not always, see the correlated random effects discussion in \citealp{chamberlain1984panel}).\footnote{The terms fixed effects and random effects are not ideal and have led to some confusion, but they are by now so widely used that we use them as well.} See for general discussions \citep{hsiao2022analysis, bell2015explaining}. This distinction between fixed and random effects was often used as an organizing principle for the panel data literature, in combination with the reliance on fixed $T$ {\it versus} large $T$ asymptotic approximations. A substantial literature was devoted to identification and inference results in settings with fixed effects leading to various forms of what \citep{neyman1948consistent} labeled the incidental parameter problem. Especially when the fixed effects entered in non-additive and non-linear ways in short (with asymptotic approximations based on fixed length) panels, with limited dependent or discrete outcomes, this led to challenging identification problems {\it e.g.,} \citep{chamberlain1980analysis, honore1992trimmed, magnac2004panel, bonhomme2012functional}. In cases where identification in fixed length settings was not feasible, the literature introduced various methods for bias-correction  (see \citep{arellano2007understanding} for a survey) or developed bounds analyses ({\it e.g.}, \citep{honore2006bounds}). More recently, these bias-reduction ideas have been extended to nonlinear two-way models ({\it e.g.}, \citealp{fernandez2016individual,fernandez2018fixed}).

The earlier econometric panel data literature paid close attention to the dynamics in the outcome process, arising from substantive questions such as the estimation of structural models for production functions and dynamic labor supply. Motivated by these questions, this literature distinguished between state dependence and unobserved heterogeneity ({\it e.g.}, \citep{heckman1981statistical, chamberlain1984panel}) and various dynamic forms of exogeneity ({\it e.g.}, weak, strong and strict exogeneity, and predeterminedness, see \citep{engle1983exogeneity, arellano1991some}). These issues have not received as much attention yet in the current literature.
The earlier literature also studied models that combined the presence of unit-fixed effects with lagged dependent variables, leading to concerns about biases of least squares estimators in short panels (the so-called Nickell bias, \cite{nickell1981biases}) and the use of instrumental variable approaches
(\citealp{nickell1981biases, arellano1991some, blundell1998initial,hahn2002asymptotically,alvarez2003time}). This literature had a huge impact on empirical work in social sciences, but the recent literature has not connected much to these issues.

In contrast, an important theme in the current literature that was not discussed as much in the earlier literature concerns the presence of general 
heterogeneity in causal effects, both over time and across units, associated with observed as well as unobserved characteristics. The recognition of the importance of heterogeneity has led to findings that previously popular estimators are sensitive to the presence of such heterogeneity and to the development of more robust alternatives. These results are related to a subset of the econometric panel data literature, {\it e.g.}, \citep{chamberlain1992efficiency,arellano2011identifying,graham2012identification,chernozhukov2013average}, which modeled heterogeneity in a way that is more in line with the current literature. We discuss this connection in detail in Section \ref{section:staggered}.

\section{Setup and Data Configurations}\label{section:configurations}

In this section, we consider three classifications of the literature. The first is based on different types of data.
The second, in terms of the relative size of the cross-section and time-series dimensions,  is familiar from the earlier literature. The third, in terms of the assignment mechanism, is original to the current literature. In the earlier literature, there was an additional classification that made a distinction that depended on the heterogeneity between cross-section units being modeled as {fixed effects} or {random effects}, {\it e.g.,} \citep{chamberlain1984panel}. This distinction plays less of a role in the current literature, although it is relevant for the design-based literature we discuss in Section \ref{section:design}.
All three classifications are helpful in understanding which specific methods may be useful and what type of asymptotic approximations for inference are credible. In addition, they allow us to place the individual papers, which often focus on particular settings, in context.

To put the following discussions into context, it is also helpful to remember that most of the recent literature has focused on the average causal effect of some intervention on the outcomes for the treated units during the periods they were treated. We do so here too, but one should keep in mind that one might be interested in an average effect beyond the study sample, or in an effect over  time periods beyond the sample period.
Later, we are more precise about the exact estimands we focus on and, in particular, how some of the assumptions, such as the absence of dynamic effects, affect both the choice of estimand and its interpretation.

\subsection{Data Types}\label{datatypes}

Although we focus in this paper mostly on the proper panel data setting where we observe outcomes for a number of units over a number of time periods, we also consider some other settings with observations at different points in time that we collectively refer to as panel data. Here we want to clarify the distinction and be precise about the notation.

\subsubsection{Panel Data.}\label{panel}

In the proper panel data case we have observations on $N$ units, indexed by $i=1,\ldots,N$, over $T$ periods, indexed by $t=1,\ldots,T$. The outcome of interest is denoted by $Y_{it}$, and the treatment is denoted by $W_{it}$, both doubly indexed by the unit and time indices. These observations may themselves consist of averages over more basic units as in the grouped repeated cross-section case from Section \ref{grcs}. We collect the outcomes and treatment assignments into two $N\times T$ matrices:
\[ \mathbf{Y}=
\left(
\begin{array}{ccccc}
	Y_{11} & Y_{12} & Y_{13}  & \dots & Y_{1T} \\
	Y_{21}  & Y_{22} & Y_{23}   & \dots & Y_{2T}  \\
	Y_{31}  & Y_{32} & Y_{33}   & \dots & Y_{3T}  \\
	\vdots   &  \vdots & \vdots &\ddots &\vdots \\
	Y_{N1}  & Y_{N2} & Y_{N3}   & \dots & Y_{NT}  \\
\end{array}
\right),\quad
 \mathbf{W}=
\left(
\begin{array}{ccccc}
	W_{11} & W_{12} & W_{13}  & \dots & W_{1T} \\
	W_{21}  & W_{22} & W_{23}   & \dots & W_{2T}  \\
	W_{31}  & W_{32} & W_{33}   & \dots & W_{3T}  \\
	\vdots   &  \vdots & \vdots &\ddots &\vdots \\
	W_{N1}  & W_{N2} & W_{N3}   & \dots & W_{NT}  \\
\end{array}
\right),
\]
with the rows corresponding to units and the columns corresponding to time periods.

We may also observe other exogenous variables, denoted by $X_{it}$ or $X_i$, depending on whether they vary over time or only by unit. Typically, we focus on a {balanced panel} where for all units $i=1,\ldots,N$ we observe outcomes for all $t=1,\ldots,T$ periods. In practice, concerns can arise from the panel being {unbalanced} either because we observe units for different lengths of time or because data is missing for some of them. We ignore both complications in the current discussion.
 
Classic examples of this proper panel setting include  \citep{ashenfelter1978estimating} with information on earnings for over 90,000 individuals for 11 years, and
\citep{abowd1989covariance} with information on wages for 1,448 individuals also for 11 years.
Another classic example is \citep{card1994minimum} with data for two periods and 399 fast-food restaurants.

\subsubsection{Grouped Repeated Cross-Section Data.}\label{grcs}

In a Grouped Repeated Cross-Section (GRCS) data setting, we have observations on $N$ units. Each unit is observed only once, in period $T_i$ for unit $i$, with the time period indexed by $i$ to account for the fact that different units may be observed at different points in time. Typically $T_i$ takes on only a few values (the repeated cross-sections) relative to the number of units, {\it e.g., } often just two or three, with many units sharing the same value for $T_i$. For some of the methods this is formally not required.  The outcome and treatment received for unit $i$ are denoted by $Y_i$ and $W_i$ respectively, both indexed just by the unit index $i$.\footnote{Some empirical studies continue to use the panel notation that includes two indices for the outcomes and treatments in the repeated cross-section case, but that is confusing because $Y_{it}$ and $Y_{it'}$ do not refer to the same unit $i$ in the repeated cross-section case.} The set of units is partitioned into two or more groups, with the group that unit $i$ belongs to denoted by $G_i\in{\cal G}=\{1,2,\ldots,G\}$.

Define the average outcome for each group/time-period pair:
\[ \overline{Y}_{gt}\equiv \left.\sum_{i=1}^N \mathbf{1}_{G_i=g,T_i=t} 
Y_i\right/ \sum_{i=1}^N \mathbf{1}_{G_i=g,T_i=t},\]
and similar for $\overline{W}_{gt}$.
If we view the $G\times T$ group averages $\overline{Y}_{gt}$, instead of the original $Y_i$, as the unit of observation, this grouped repeated cross-section setting is just like a panel as in Section \ref{panel}, immediately allowing for methods that require repeated observations on the same unit. This was  pointed out in
\cite{deaton1985panel, wooldridge2010econometric}. Many methods in the GRCS literature do not use the data beyond the group/time averages, and so the formal distinction between the grouped repeated cross-section and proper panel case becomes moot. However, in practice, empirical applications with grouped repeated cross-section data have typically many fewer groups than proper panel data have units, sometimes as few as two or three, limiting the scope for high-dimensional parametric models and raising concerns about the applicability of large-$N$ asymptotics.

In a seminal application of DID estimation with repeated cross-section data, with two groups and two periods, \citep{meyer1995workers}, the units are individuals getting injured on the job, and we observe individuals getting injured at most once. 
The time periods correspond to the  year the individuals are injured, with data available for two years. Similarly, in \citep{eissa1996labor} the units are different taxpayers in two different years, with the number of groups again equal to two.
The case with more than two groups is studied in \citep{Bertrand2004did}, and in countless other studies, often with the groups corresponding to states, and the treatment regulations implemented at the state level.

\subsubsection{Row and Column Exchangeable Data}\label{twdf}

One data type that has not received as much attention as either panel or repeated cross-section data corresponds to what we refer to as row-column exchangeable data (RCED), 
(\citep{aldous1981representations, lynch1984canonical}).  Like proper panel data, these data are doubly indexed, with outcomes denoted by $Y_{ij}$, $i=1,\ldots,N$, $j=1,\ldots,J$. The difference with panel data is that there is no ordering for the second index (time in the proper panel case). An example of such a data type is supermarket shopping data, where we observe expenditures on item $j$ for shopper $i$, or data from a rideshare company, where we observe outcomes for trips involving customer $i$ and driver $j$, or a customer/product setting for an online retailer (\citealp{abadie2024doubly}). Although this is not a particularly common data configuration, it is useful to contrast  it explicitly with proper panel  and cross-section data.  Proper panel data differ in two aspects from cross-section data: the double indexing and the time ordering: the RCED setting is in-between the cross-section and proper panel case, with the double indexing but no time ordering.

In this case, where the second index is not time, it is natural to model both units $i=1,\ldots,N$ and $j=1,\ldots,J$ as exchangeable, whereas with proper panel data, the exchangeability of the time periods is typically implausible. It is interesting to note that 
many,  but not all, methods ostensibly developed for use with panel data are also applicable in this RCED setting. For example, TFWE methods, factor models, and many SC estimators, all discussed in more detail below, can be used with such data. The fact that those methods can be used in the RCED setting directly means that such estimators do not place any value on knowledge of the time series ordering of the data. If {\it ex ante} one believes such information is valuable, one may wish to use methods that exploit it.

A related but even more general data type involves RCED with repeated observations. An example of such a data frame is a panel of matched employer-employee data (\textit{e.g.}, \citealp{abowd1999high, card2022industry}). See \citep{bonhomme2020econometric} for a recent survey of the relevant methods.

\subsection{Shapes of Data Frames}

Our second classification of the panel data literature is organized by the shape of the data
frame. This is not an exact classification, and which category a particular data set fits,
and which methods are appropriate, in part depends on the magnitude of the cross-
section and time-series correlations and not just on the magnitude of $N$ and $T$. Nevertheless, it is useful to reflect on the relative
magnitude of the cross-section and time-series dimensions as it has implications for the
properties of statistical methods for the analysis of such data. In particular, it often
motivates the choice of asymptotic approximations based on large $N$ and fixed $T$, or
large $N$ and large $T$.

\subsubsection{Thin Data Frames: Many Units, Few Time Periods ($N\gg T$)}\label{manyunits}

Much of the traditional panel data case considers the setting where the number of cross-section units is large relative to the number of time periods:
\[
\begin{array}{r}\by^\tall=\\ 
(N\gg T)
\end{array}
\left(
\begin{array}{ccccccc}
	Y_{11} & Y_{12} & Y_{13}   \\
	Y_{21}  & Y_{22} & Y_{23}   \\
	Y_{31}  & Y_{32} &  Y_{33}    \\
		Y_{41}  & Y_{42} & Y_{43}   \\
		Y_{51}  & Y_{52} & Y_{53}  \\	
	Y_{61}  & Y_{62} & Y_{63}    \\
	\vdots   &  \vdots & \vdots \\
	Y_{N1}  & Y_{N2} &  Y_{N3}    \\
\end{array}
\right)\]
This is a common setting when the units are individuals and it is challenging or expensive to get repeated observations for many periods for the same individual. The PSID and NLS panel data fit this setting, with often thousands of units. In this case inferential methods often rely on asymptotic approximations based on large $N$ for fixed $T$. Incidental parameter problems of the type considered by \citep{neyman1948consistent} are particularly relevant (see \citep{lancaster2000incidental} for a modern discussion). Specifically, if there are unit-specific parameters, {\it e.g.,} fixed effects, it is not possible to estimate those parameters consistently. This does not necessarily imply that one cannot estimate the target parameters consistently, and the traditional literature developed many procedures that allowed for the elimination of these fixed effects, even if they enter nonlinearly, {\it e.g.,} \citep{honore1992trimmed, chamberlain2010binary,bonhomme2012functional}. However, the fact that the time series dimension is small or modest does mean that random effect assumptions are potentially powerful because they place a stochastic structure on the individual components so that these individual components can be integrated out.

\subsubsection{Fat Data Frames: Few Units, Many Time Periods ($N\ll T$)}
The second setting is one where the number of time periods is large relative to the number of cross-section units:
\[
 \begin{array}{r}\by^\fat=\\ 
(N\ll T)
\end{array}
\left(
\begin{array}{cccccccccccc}
	Y_{11} & Y_{12} & Y_{13}   & Y_{14} & Y_{15} & Y_{16} & Y_{17} & Y_{18} &\dots & Y_{1T} \\
	Y_{21} & Y_{22}  & Y_{23}  & Y_{24} & Y_{25} & Y_{26} & Y_{27} & Y_{28}   & \dots & Y_{2T}  \\
	Y_{31}  & Y_{32} & Y_{33}  & Y_{34} & Y_{35}   & Y_{36} & Y_{37} & Y_{38} & \dots & Y_{3T}  \\
	 Y_{41}  & Y_{42} & Y_{43}  & Y_{44}  & Y_{45}& Y_{46}  & Y_{47} & Y_{48}   & \dots & Y_{4T}  \\
\end{array}
\right)
\]
This setting is more common when the cross-section units are aggregates, {\it e.g.,} states or countries, for which we have observations over many time periods, say output measures for quarters, or unemployment rates per month.

This setting is closely related to the traditional time series literature, but the insights from that literature have not always been fully appreciated in the modern causal panel literature. There are some exceptions that take more of a time-series approach to this type of panel data, {\it e.g.,} \citep{brodersen2015inferring, ben2023estimating}. The work on inference using conformal methods is also in this spirit, {\it e.g.,} \citep{chernozhukov2021exact}.

\subsubsection{Square Data Frames: Comparable Number of Units and Time Periods: ($N\approx T$)}
In the third case  the number of time periods and cross-section units is roughly comparable:
\[ \begin{array}{r}\by^\squ=\\ 
(N\approx T)
\end{array}
\left(
\begin{array}{ccccccc}
	Y_{11} & Y_{12} & Y_{13}  & \dots & Y_{1T} \\
	Y_{21}  & Y_{22} & Y_{23}   & \dots & Y_{2T}  \\
	Y_{31}  & Y_{32} & Y_{33}   & \dots & Y_{3T}  \\
	\vdots   &  \vdots & \vdots &\ddots &\vdots \\
	Y_{N1}  & Y_{N2} & Y_{N3}   & \dots & Y_{NT}  \\
\end{array}
\right)\]
A common example is where the units are states and the time periods are years or quarters. We may have observations on 50 states for 30 years or 80 quarters. This is a particularly challenging case but, at the same time, increasingly common in practice. Many empirical studies using DID/TWFE, SC, or related estimators fit into this setting.

Whether in this case asymptotic approximations based on large $N$ and fixed $T$, or large $N$ and large $T$, or neither, are appropriate is not always obvious. 
Simply looking at the magnitudes of the time series and cross-section dimension itself is not sufficient to make that determination because the appropriate approximations also depend on the magnitude of cross-section and time-series correlations. There is an important lesson in this regard in the weak instrument literature. In the influential Angrist-Krueger analysis of the returns to schooling \citep{angristkrueger1991}, the authors report results based on over 300,000 units and 180 instruments. Because of the relative magnitude of the number of units and instruments, one might have expected that asymptotic approximations based on a fixed number of instruments and an increasing number of units would be appropriate. Nevertheless, it turned out that the Bekker asymptotic approximation, developed by \citep{bekker1994alternative} and based on letting the number of instruments increase proportionally to the number of units, is substantially more accurate because of the weak correlation between the instruments and the endogenous regressor (years of education in the Angrist-Krueger study).

The earlier
econometric panel data literature discusses the tradeoffs between various asymptotic approximations for the analysis of dynamic linear models, \textit{e.g.,} see  \cite{hahn2002asymptotically,alvarez2003time}. In dynamic models the fixed effect estimator is inconsistent in short panels. Alternative estimators have been proposed using lagged outcomes as instruments (\cite{arellano1991some, blundell1998initial}). As the panel becomes longer, the number of instruments grows. However, the more distant lags are often only weakly correlated with the endogenous regressors, leading to many weak instruments problems. One important aspect of the panel data analysis is that the fixed effect estimator is consistent in the large-$T$ limit but not in the fixed $T$ setting.

\subsection{Assignment Mechanisms}

The third classification for panel data methods we consider is based on features of the assignment process for the treatment. As in the classification based on the relative magnitudes of the components of the data frame, features of the assignment process are important for determining which statistical methods and which asymptotic approximations are reasonable. This classification is not present in the earlier panel data literature but features prominently in the current literature. This reflects the more explicit focus on causal effects in general in the econometric literature of the last three decades. It should be noted that this classifications is not so much based on assumptions such as endogeneity or exogeneity, as it is about facts, regarding the assignment process.

One feature that is common to many applications of the methods is that the fraction treated unit/time-period pairs is small. This has two implications. First, the focus is typically on the average effect on the treated unit/time-period pairs. This may be for substantive reasons but it is also motivated by the fact that if the fraction of treated pairs is small, the precision of estimates for the overall average effect will be considerably lower than the precision of estimates for the average effect for the treated pairs.
Second, building statistical models for the treated outcomes will be of low value, as such models will not increase the precision of standard estimators. Thus, important modeling questions are about the control outcomes.

\subsubsection{The General Case}\label{general}

In the most general case the treatment may vary both across units and over time, with units switching in and out of the treatment group:
\begin{equation}\label{eq:general}
\begin{array}{r}\bw^\gen=\\ 
({\rm general})
\end{array}\left(
\begin{array}{cccccccc}
	{\ttock} & {\ttock} & \ttick  & \ttick & \dots & {\ttock} \\
	\ttick  & \ttick & {\ttock} & \ttick   & \dots & \ttick  \\
	{\ttock}  & \ttick & {\ttock}  & \ttock  & \dots & \ttick  \\
	{\ttock}  & \ttick & \ttick & {\ttock}   & \dots & \ttick  \\
	\vdots   &  \vdots  &  \vdots& \vdots &\ddots &\vdots \\
	{\ttock}  & \ttick & {\ttock}   & \ttick & \dots & \ttick  \\
\end{array}
\right)
\end{equation}
With this type of data, we can use variation of the treatment within units and variation of the treatment within time periods to identify causal effects. Especially in settings without dynamic effects, the presence of both types of variation may improve the credibility of estimators for causal effects. This setting is particularly relevant for the RCED configurations but it is less common in proper panel data settings. Some examples include marketing settings with the units corresponding to products and the treatment corresponding to promotions or discounts.

In this setting, assumptions about the absence or presence of dynamic treatment effects are particularly important. In applications where dynamic treatment effects are present, many commonly used methods assuming their absence lead to difficult-to-interpret results.

\subsubsection{Single Treated Period}

One important special case arises when a substantial number of units is treated, but these units are only treated in the last period. 
\[
\begin{array}{r}\bw^\last=\\ 
({\rm last\ period})
\end{array}\left(
\begin{array}{cccccccc}
	 \ttick &  \ttick & \ttick  & \ttick & \dots & \ttick \\
	\ttick  & \ttick &  \ttick & \ttick   & \dots & \ttick  \\
	 \ttick  & \ttick &  \ttick  & \ttick  & \dots & \ttock  \\
	 \ttick  & \ttick & \ttick &  \ttick   & \dots &  {\ttock}  \\
	\vdots   &  \vdots  &  \vdots& \vdots &\ddots &\vdots \\
	 \ttick  & \ttick &  \ttick   & \ttick & \dots & {{\ttock}}  \\
\end{array}
\right).
\]
In settings where the number of time periods is relatively small, this case is often analyzed as a cross-section problem. The lagged outcomes are simply used as exogenous covariates or pre-treatment variables that should be adjusted for in treatment-control comparisons based on an unconfoundedness assumption (\citealp{rosenbaum1983central}). A classic example in the economics literature is the Lalonde-Dehejia-Wabha data originally collected in \citep{lalonde1986evaluating} with the data set now commonly used constructed and analyzed in \citep{dehejiawahba}.
This data set has served as a valuable playground for assessing new methodological advances in the literature on unconfoundedness.
In that case, there are three periods of outcome data (earnings) but only one post-treatment outcome. The original study \citep{lalonde1986evaluating} reported results for a variety of models, including some two-way-fixed-effect regressions. Much of the subsequent literature since \citep{dehejiawahba, dehejia2002propensity} has focused more narrowly on methods relying on unconfoundedness, sometimes in combination with functional form assumptions. See \citep{imbens2024lalonde} for a recent discussion. Asymptotics are typically, and appropriately so, based on large $N$ and fixed $T$.

Given that the treatment is observed only in the last period, the presence of dynamic effects is not testable, and dynamics do not really matter in the sense that their presence only leads to a minor change in the interpretation of the estimand, typically the average effect for the treated units and time periods. Because of the shortness of the panel, these are obviously short-term effects, with little evidence regarding the long-term impacts of the interventions.

\subsubsection{Single Treated Unit}
Another key setting is that with a single treated unit, treated in multiple periods.
\[ 
\begin{array}{r}\bw^\single=\\ 
(\text{single unit})
\end{array}\left(
\begin{array}{cccccccc}
	\ttick & \ttick & \ttick  & \ttick & \dots & \ttick \\
	\ttick  & \ttick & \ttick & \ttick   & \dots & \ttick  \\
	\ttick  & \ttick & \ttick  & \ttick  & \dots & \ttick  \\
	\ttick  & \ttick & \ttick & \ttick   & \dots & \ttick  \\
	\vdots   &  \vdots  &  \vdots& \vdots &\ddots &\vdots \\
	\ttick  & \ttick & \ttock   & {\ttock} & \dots & {\ttock}  \\
\end{array}
\right).\]  
This setting is prominent in the original applications of the {synthetic control} literature: \citep{abadie2003,abadie2010synthetic, abadie2019using}. This literature has exploded in terms of applications and theoretical work in the last twenty years. Here the number of time periods is typically too small to rely credibly on large $T$ asymptotics, creating challenges for inference that are not entirely resolved. Large $N$ asymptotics creates its owm, different, challenges, stemming from the lack of multiple treated units.

\subsubsection{Single Treated Unit and Single  Treated Period}
An extreme case is that with only a single unit ever treated, and this unit only treated in a single period, typically the last period:
\[
\begin{array}{r}\bw^{\rm one}=\\ 
(\text{single unit/period})
\end{array}\left(
\begin{array}{cccccccc}
	 \ttick &  \ttick & \ttick  & \ttick & \dots & \ttick \\
	\ttick  & \ttick &  \ttick & \ttick   & \dots & \ttick  \\
	 \ttick  & \ttick &  \ttick  & \ttick  & \dots & \ttick  \\
	 \ttick  & \ttick & \ttick &  \ttick   & \dots &\ttick  \\
	\vdots   &  \vdots  &  \vdots& \vdots &\ddots &\vdots \\
	 \ttick  & \ttick &  \ttick   & \ttick & \dots & {{\ttock}}  \\
\end{array}
\right).
\]
This is a challenging setting for inference: we cannot rely on large sample approximations for the average outcomes for the treated unit/periods because there is only a single treated unit/period pair. 
Instead of focusing on population parameters it is natural here to focus on the effect for the single treated/time-period pair and construct prediction intervals.
Because it is a special case of both the single treated period and the single treated unit case it is conceptually important for comparing estimation methods popular for those settings.

\subsubsection{Block Assignment}\label{block}

Another important case in practice is that with block assignment, where a subset of units is treated  every period after a common starting date:
\[ 
\begin{array}{r}\bw^\block=\\ 
({\rm block})
\end{array}\left(
\begin{array}{ccccccccc}
	\ttick & \ttick & \ttick  & \ttick & \dots & \ttick  & \ttick\\
	\ttick  & \ttick & \ttick & \ttick   & \dots & \ttick  & \ttick \\
	\ttick  & \ttick & \ttick  & \ttick  & \dots & \ttick  & \ttick \\
	\ttick  & \ttick & \ttick & {\ttock}   & \dots & {\ttock}  & {\ttock} \\
	\vdots   &  \vdots  &  \vdots& \vdots &\ddots &\vdots &\vdots\\
	\ttick  & \ttick & \ttick   & {\ttock} & \dots & {\ttock} & {\ttock}  \\
\end{array}
\right)\]
This assignment matrix is the basis of the simulations reported in 
\citep{Bertrand2004did} and \citep{ arkhangelsky2021synthetic}. In this case there is typically a sufficient number of treated unit/time-period pairs to allow for reasonable approximations to be based on that number being large. 

Here the presence of dynamic effects changes the interpretation of the average effect for the treated. The average effect for the treated now becomes an average over short and medium term effects during different periods. There is limited ability to separate out heterogeneity across calendar time and dynamic effects because, in any given time period, there are only treated units with an equal number of treated periods in their past.

\subsubsection{Staggered Adoption}\label{staggered}

The recent DID/TWFE literature has focused on the staggered adoption case where units remain in the treatment group once they adopt the treatment, but they vary in the time at which they adopt the treatment. Some may adopt early, while others adopt  later:
\[
\begin{array}{r}{\bw^\stag=} \\
{(\text{staggered  adoption})}
\end{array}\left(
\begin{array}{ccccccccc}
	\ttick &  \ttick & \ttick  & \ttick & \dots & \ttick & \ttick \\
	\ttick  & \ttick &  \ttick & \ttick   & \dots  & \ttick& {\ttock}  \\
	\ttick  & \ttick &  \ttick  & \ttick  & \dots  & \ttock& {\ttock}  \\
	\ttick  & \ttick & \ttick &  {\ttock}   & \dots & \ttock & {\ttock}  \\
	\vdots   &  \vdots  &  \vdots& \vdots &\ddots &\vdots &\vdots \\
	\ttick  & {\ttock} &  {\ttock}   & {\ttock} & \dots  & \ttock& {\ttock}  \\
\end{array}
\right)
\]
This case is also referred to as the {absorbing treatment} setting. Clearly, this setting leads to much richer information about the possible presence of dynamic effects, with the ability, under some assumptions, to separate dynamic effects from heterogeneity across calendar time.

A second issue is whether, for units adopting in period $t$, the best controls are units adopting in period $t+1$, or later, or possibly the units never adopting the treatment (\citealp{callaway2020difference}).

\subsubsection{Event Study Designs}\label{event_study}

A closely related design is the event-study design, where units are exposed to the treatment in at most one period.
\[
\begin{array}{r}{\bw^{\mathrm{event}}=} \\
{(\text{event study})}
\end{array}\left(
\begin{array}{ccccccccc}
	\ttick &  \ttick & \ttick  & \ttick & \dots & \ttick & \ttick \\
	\ttick  & \ttick &  \ttick & \ttick   & \dots  & \ttick& {\ttock}  \\
 \ttock  & \ttick &  \ttick & \ttick   & \dots  & \ttick& {\ttick}  \\
	\ttick  & \ttick &  \ttock  & \ttick  & \dots  & \ttick& {\ttick}  \\
	\ttick  & \ttick & \ttock &  {\ttick}   & \dots & \ttick & {\ttick}  \\
	\vdots   &  \vdots  &  \vdots& \vdots &\ddots &\vdots &\vdots \\
	\ttick  & {\ttick} &  {\ttick}   & {\ttock} & \dots  & \ttick& {\ttick}  \\
\end{array}
\right)
\]
In this setting there are often dynamic effects of the treatment past the time of initial treatment. If these effects are identical to the initial effect, the analysis can end up being very similar to that of staggered adoption designs. 
Canonical applications include some in finance, {\it e.g.,} \cite{fama1969adjustment}.

\subsubsection{Clustered Assignment}

Finally, in many applications, units are grouped together in clusters, with units within the same clusters always assigned to the same treatment. The example below has $C$ clusters, with a subset of the clusters assigned to the treatment from a common period onwards in a block assignment structure. 
\[ 
\begin{array}{r}\bw^\cluster=\\ 
({\rm cluster/block})
\end{array}\left(
\begin{array}{ccccccccccc}
& &   & & &   &  & &\textrm{cluster}\\
 & &   & & &   &  & &\downarrow\\
	\ttick & \ttick & \ttick  & \ttick & \dots & \ttick  & \ttick & & 1\\
	\ttick & \ttick & \ttick  & \ttick & \dots & \ttick  & \ttick & & 1\\
	\ttick  & \ttick & \ttick & \ttick   & \dots & \ttick  & \ttick& & 1 \\
	\ttick  & \ttick & \ttick  & \ttick  & \dots & \ttick  & \ttick& & 2 \\
	\ttick  & \ttick & \ttick  & \ttick  & \dots & \ttick  & \ttick & & 2\\
	\ttick  & \ttick & \ttick & {\ttock}   & \dots & {\ttock}  & {\ttock} & & 3\\
	\vdots   &  \vdots  &  \vdots& \vdots &\ddots &\vdots &\vdots & & \vdots\\
	\ttick  & \ttick & \ttick   & {\ttock} & \dots & {\ttock} & {\ttock} & & C  \\
 \ttick  & \ttick & \ttick   & {\ttock} & \dots & {\ttock} & {\ttock}  & & C \\
\end{array}
\right)\]
The clustering creates particular complications for inference, whether it is in the block assignment case, or other settings, in particular because often there are relatively few clusters. It also creates challenges for  estimation if there are cluster components to the outcomes.

\section{Potential Outcomes, General Assumptions, and Estimands}\label{notation}

In this section we collect in a single section the notation that allows us to cover various parts of the literature. We focus on the proper panel data case with $N$ units and $T$ periods. We use the potential outcome notation (see \citealp{rubin1974estimating, imbens2015causal}). We also discuss basic estimands that have been the focus of this literature and some of the maintained assumptions. 

Let $\ubww$ denote the full $T$-component column vector of treatment assignments,
\[ \ubww\equiv (w_1,\ldots,w_T)^\top,\]
and $\ubw_i$ the vector of treatment values for unit $i$. Let
 $\ubww^t$ the $t$-component column vector of treatment assignments up to time $t$:
\[ \ubww^t\equiv (w_1,\ldots,w_t)^\top,\]
so that $\ubww^T=\ubww$, and similar for $\ubw^t_i$.
In general we can index the potential outcomes for unit $i$ in period $t$ by the full $T$-component vector of assignments $\ubww$:
\[ Y_{it}(\ubww).\]
Even this notation already makes a key assumption, the Stable Unit Treatment Value Assumption, or SUTVA (see \citealp{rubin1978bayesian, imbens2015causal}). SUTVA requires that there is no interference or spillovers between units. This is a strong assumption, and in many applications, it may be violated. There has been little attention paid to models allowing for such interference in the recent causal panel data literature to date, although there is extensive literature on interference in cross-section settings, {\it e.g.}, in clustering settings (\citealp{hudgens, manski1993}), in network settings (\citealp{leung2023network, auerbach2022identification}), and in the general case (\citealp{aronow2017estimating}). 

In applications where the spillover effects are only present within certain groups, e{\it .g.,} clusters, or economic markets, and the treatment is assigned at the same level, one can justify SUTVA by aggregating the individual data to the cluster or group level. In this case, the potential outcome introduced above would correspond to the aggregated data. This is directly connected to our discussion of Grouped Repeated Cross Section (GRCS) data in Section \ref{grcs}. Of course, the aggregation changes the interpretation of the causal effects, which would now incorporate both direct and spillover effects.  

Without further restrictions, our setup describes for each unit and each time period $2^T$ potential outcomes, as a function of multi-valued treatment $\ubww$. As a result we can define for every period $t$  unit-level treatment effects for every pair of assignment vectors $\ubww$ and $\ubww'$:
\begin{equation}\label{eq:estimand_general}
\tau^{\ubww,\ubww'}_{it}\equiv Y_{it}(\ubww') -Y_{it}(\ubww),
    \end{equation}
with the corresponding population average effect defined as
\[\tau^{\ubww,\ubww'}_{t}\equiv\mme\left[ Y_{it}(\ubww') -Y_{it}(\ubww)\right].
\]
These unit-level and average causal effects are the basic building blocks of many of the estimands considered in the literature. 
Note that we implicitly assume there is a large population over which we can take the expectation. Part of the literature has focused on finite sample issues using a design perspective. See Section  \ref{section:design} for more discussion on this.

If we are only interested in average causal effects of the form $\tau^{\ubww,\ubww'}_{t}$, 
then we have, in essence, a problem similar to the cross-sectional version of the problem of estimating average causal effects. One approach would be to analyze such problems using standard methods for multi-valued treatments under unconfoundedness, {\it e.g.}, \citep{imbens2000}.  Here this would require comparing outcomes in period $t$ for units with treatment vectors $\ubww$ and $\ubww'$.

If we have completely random assignment, all average causal efects of the type $\tau^{\ubww,\ubww'}_{t}$ are identified, given sufficient variation in the treatment paths. That also means that we can identify in this setting dynamic treatment effects. For example, in the two-period case
\[ \tau^{(1,1),(0,1)}_2,\]
is the average effect in the second period of being exposed to the sequence $(1,1)$ rather than the sequence $(0,1)$, so it measures the dynamic effect of a sequence of treatments on period 2 outcomes, being exposed to the treatment in both periods versus being exposed only in the second period.

A key challenge is that there are many, 
 $2^{T-1}\times(2^T-1)$ to be precise, distinct average effects of the form $\tau^{\ubww,\ubww'}_{t}$. 
Even with $T=2$  there are already six different average causal effects, and with $T$ larger, this number quickly increases. This means that in practice we need to limit or focus on summary measures of all these causal effects, {\it e.g.,} averages over effects at different times. Typically we also put additional structure on these causal effects in the form of cross-temporal restrictions on the potential outcomes $Y_{it}(\ubww)$. That enables us to give comparisons of outcomes from different time periods a causal interpretation. See \citep{chamberlain1984panel} for a discussion of this in the case of linear models. Note that without additional restrictions, all the average treatment effects  $\tau^{\ubww,\ubww'}_{t}$ are just-identified, so any additional assumptions will typically imply testable restrictions.

The first restriction that we consider is the commonly made no-anticipation assumption, {\it e.g.}, \citep{athey2021design, callaway2020difference, sun2020estimating}. This requires that potential outcomes do not depend on future treatments.
\begin{assumption}{\sc (No Anticipation)}\label{as:no_an}
The potential outcomes satisfy
\[Y_{it}(\ubww)=Y_{it}(\ubww'), \]
for all $i$, and for all combinations of $t,$ $\ubww$ and $\ubww'$ such that
$\ubww^t=\ubww^{\prime t}$.
\end{assumption}
With experimental data, and sufficient variation in treatment paths, this assumption is testable. To do so we can compare outcomes in period $t$ for units with the same treatment path up to and including $t$, but whose treatment paths diverge in the future, that is, after period $t$. The average difference between such average outcomes should be zero in expectation under the no-anticipation assumption.

This substantive assumption is appealing in situations where units are not active decision-makers but rather passive recipients of the treatment. In such cases, the no-anticipation assumption can, in principle, be guaranteed by design. If random units are assigned treatment each period, or, in the staggered adoption case, if the adoption date is randomly assigned, potential outcomes cannot vary with the future random assignment. Of course, in observational studies, the assumption need not hold. In many applications, treatments are state-level regulations that are known to be coming prior to the time they formally take effect. One remedy for this problem is to allow for limited anticipation, assuming the treatment can be anticipated for a fixed number of periods, as in \citep{callaway2020difference}. Algorithmically, this amounts to redefining $\ubww$ by shifting it by the fixed number of periods.

At the same time, there are numerous economic applications where units are involved in an active decision-making process. Units can make decisions about variables for which $\ubww$ is an important input, and beliefs about future treatment paths would affect those decisions. For example, current taxes and beliefs about future tax changes can be important determinants of current consumption. In this case, researchers first need to address a key conceptual issue. The premise of the potential outcome framework is that it describes the exhaustive set of counterfactual outcomes that can be realized in an experiment where the researcher controls the assignment of $\ubww$. However, if the units are making decisions in environments with uncertainty, then they can change their behavior in response to different distributions of the future treatment paths, in line with Lucas's critique (\citealp{lucas1976econometric}). As a result, one cannot express potential outcomes as functions of $\ubww$ only but also needs to view them as functions of the experimental design itself, {\it i.e.}, the known or anticipated distribution of $\ubww$. 

One solution to this problem is to define potential outcomes for a given randomized experimental design. 
Assumption \ref{as:no_an} then becomes innocuous because the beliefs about the future treatment paths are incorporated in the definition of the potential outcomes, and the actual values are by construction unknown.  This does, however, change the interpretation of the causal effects. This issue is well understood in macroeconomic literature, which emphasizes the distinction between the effect of a surprise deviation
from a given policy rule versus the effect of a permanent change in the policy rule itself. While the former quantity can be learned using various quasi-experimental strategies (see \citep{nakamura2018identification} for a discussion), the identification of the latter typically relies on an economic model (see, however, \citealp{McKay2023}). Causal panel data literature could benefit from explicitly incorporating these ideas. See \citep{abbring2007econometric} for a related discussion and additional references.  

The situation becomes considerably more complicated in observational studies, where one cannot directly control the information about the future treatment paths available to the units. Nevertheless, in some applications, the researchers directly observe the arrival of such information. In this case, to make the Assumption \ref{as:no_an} plausible, one needs to guarantee that different units face the same informational environment. Failure to do so is akin to comparing outcomes across units participating in experiments with different designs. \citep{abbring2003nonparametric} and \citep{abbring2007econometric} show that many economic applications have data that would allow researchers to measure the information inflow and discuss how to adjust for the differences across units in this inflow to ensure that Assumption \ref{as:no_an} holds.

The no anticipation assumption reduces the total number of potential treatment effects from $ 2^{T-1}\times(2^T-1)$ to $(\sum_{t=1}^T 2^{t-1})(\sum_{t=1}^T 2^t-1)$. The basic building blocks, unit-period specific treatment effects, are now of the type
\begin{equation}\label{eq:estimand_general_1}
\tau^{\ubww^t,\ubww^{t^{\prime}}}_{it}\equiv Y_{it}(\ubww^{t^\prime}) -Y_{it}(\ubww^t),
\end{equation}
with the potential outcomes for period $t$ indexed by treatments up to period $t$ only.

This current structure still allows us to distinguish between static treatment effects, {\it i.e.,} $\tau^{(\ubww^{t-1},0),(\ubww^{t-1},1)}_{it}$, which measures the response of current outcome to the current treatment, holding the past ones fixed, and dynamic ones, i.e., $\tau^{(\ubww^{t-1},w^t),({\ubww^{\prime}}^{t-1},w^t)}_{it}$, which does the opposite. In the earlier panel data literature, the dynamic effects were explicitly modeled by putting a particular structure on them, but in principle, one can identify them without imposing additional restrictions on the potential outcomes given assumptions on the assignment mechanism, such as random assignment, {\it e.g., } \citep{bojinov2021panel}. There is also a large literature in biostatistics on dynamic models that is relevant for these problems, {\it e.g., } \citep{robins2000marginal, murphy2003optimal}.

A stronger assumption is that the potential outcomes only depend on the contemporaneous assignment, ruling out dynamic effects of any type.
\begin{assumption}{\sc (No Dynamic / Carry-over Effects)}
    The potential outcomes satisfy
\[Y_{it}(\ubww)=Y_{it}(\ubww'), \]
for all  $i$ and for all combinations of $t$, $\ubww$ and $\ubww'$ such that
$w_{it}=w_{it}'$.
\end{assumption}
This is \emph{not} a design assumption that can be guaranteed by randomization in a suitably designed experiment. It restricts the treatment effects and, thus, the potential outcomes for the post-treatment periods. Like the no-anticipation assumption it has testable restrictions given the random assignment of the treatment and sufficient variation in the treatment paths.
Note that it does \emph{not} restrict the time path of the potential outcomes in the absence of any treatment, $Y_{it}(\mathbf{0})$, where $\mathbf{0}$ is the vector with all elements equal to zero. In fact, these outcomes can exhibit arbitrary correlations in the sequence of potential outcomes $Y_{it}(\underline{w})$ for any given $\underline{w}$. 

If we are willing to make the no-dynamic effects assumption, we can write the potential outcomes, with some abuse of notation, as $Y_{it}(0)$ and $Y_{it}(1)$ with a scalar argument. In this case, the total number of treatment effects for each unit is greatly reduced to $T$ (one per period), and we can simplify them  to
\begin{equation}\label{eq:estimand_general_2}
\tau_{it} \equiv Y_{it}(1) - Y_{it}(0),
\end{equation}
where $\tau_{it}$ has no superscripts because there are only two possible arguments of the potential outcomes, $w\in\{0,1\}$.

So far, we have discussed assumptions on potential outcomes themselves. A conceptually different assumption is that of absorbing treatments, that is where the assignment mechanism corresponds to staggered adoption.
\begin{assumption}{\sc (Staggered Adoption)}
\[ W_{it}\geq W_{it-1}\quad\forall t=2,\ldots,T.\]    
\end{assumption}
Defining the adoption date $A_i$ as the date of the first treatment, $A_i\equiv T+1-\sum_{t=1}^T W_{it}$ for units that are treated in the sample, and $A_i \equiv \infty$ for never-treated ones.
In the staggered adoption case, we can write the potential outcomes, again with some abuse of notation, in terms of the adoption date, $Y_{it}(a)$, for $a=1,\ldots,T, \infty$, and the realized outcome as $Y_{it}=Y_{it}(A_i)$.

There are two cases that are sometimes viewed as staggered adoption designs but that are  different in substance although not always in terms of analyses. First, there may be interventions that are adopted and remain in place. States or other administrative units adopt new regulations at different times. For example, states adopted speed limits or minimum drinking ages at different times (\citealp{ashenfelter2004using}), and counties adopted enhanced 911 policies at different times (\citealp{athey2002impact}). 
These staggered adoption designs were introduced in Section \ref{staggered}.
Second, there may be one-time interventions that have a long-term or even permanent impact. We refer to such settings, introduced in Section \ref{event_study} as event studies. In that case, the post-intervention period effects would be dynamic effects.

Given staggered adoption but absent the no-anticipation  and no-dynamics assumptions, we can write the building blocks as
\begin{equation}\label{eq:athey_imbens}
\tau^{a,a'}_{it}\equiv Y_{it}(a') -Y_{it}(a),
    \end{equation}
with the corresponding population average
\[ \tau^{a,a'}_{t}\equiv \mme\left[ Y_{it}(a') -Y_{it}(a)\right].
\]
We also introduce notation for the average for subpopulations defined by the adoption date:
\[ \tau^{a,a'}_{t|a''}\equiv\mme\left[Y_{it}(a') -Y_{it}(a)|A_i = a''\right].\] 
Compared to previously defined estimands, this one explicitly depends on the details of the assignment process, which determines which units adopt the treatment and when they do so. This estimand is conceptually similar to the average effect on the treated in cross-sectional settings, with the important difference that selection now operates over two dimensions: units and periods. As in the cross-sectional setting, this matters for interpretation in observational studies, in which the researcher does not control the assignment process.

In the two-period case where all units are exposed to the control treatment in the initial treatment, the estimand $ \tau^{0,1}_{t|1}$, the average effect of the treatment in the second period for those who adopt in the second period is very much like the average effect for the treated, In settings with more variation in the adoption date there are many such average effects, depending on when the units adopted, and which period we are measuring the effect in.

\section{Two-Way-Fixed-Effect  and Difference-In-Differences Estimators}\label{section:twe}

In this section we give a brief introduction to conventional Difference-In-Differences (DID) 
or Two-Way-Fixed-Effect (TWFE) estimation. The discussion is framed in terms of the potential outcomes framework from the modern causal inference literature, but otherwise it is largely standard and following textbook discussions. 
For other recent surveys of this literature see \citep{chiu2023and, de2023two, roth2023s}.

\subsection{The Two-Way-Fixed-Effect Characterization}

We start with the Two-Way-Fixed-Effect (TWFE) specification in a proper panel setting with no anticipation and no dynamics and parallel trends or constant treatment effects. Traditionally this specification often motivates the DID estimator. 
\begin{assumption}\label{ass:twfe}{\sc (The Two-Way-Fixed-Effect Model)}
    The control outcome $Y_{it}(0)$ satisfies a two-way-fixed-effect structure:
\begin{equation}\label{eq:twfe} Y_{it}(0)=\alpha_i+\beta_t+\varepsilon_{it}.\end{equation}
The unobserved component $\varepsilon_{it}$ is (mean-)independent of the treatment assignment $W_{it}$.
\end{assumption}
\begin{assumption}\label{ass:constant}{\sc (Parallel Trends Assumption)}
 The potential outcomes satisfy
\[ Y_{it}(1)=Y_{it}(0)+\tau\quad\forall (i,t).\]
\end{assumption}
The combination of these two assumptions leads to a model for the realized outcome, defined as $Y_{it}\equiv W_{it} Y_{it}(1)+(1-W_{it}) Y_{it}(0)$, 
\begin{equation}\label{eq:twfe_obs} Y_{it}=\alpha_i+\beta_t+\tau W_{it}+\varepsilon_{it}.\end{equation}
We can estimate the parameters of this model by least squares:
\begin{equation}\label{estimator:twfe}
(\hat\tau^\twfe,\hat\alpha,\hat\beta)=\arg\min_{\tau,\alpha,\beta}
\sum_{i=1}^N\sum_{t=1}^T \left( Y_{it}-\alpha_i-\beta_t-\tau W_{it}\right)^2.
\end{equation}
Here we need to impose one restriction on the $\alpha_i$ or $\beta_t$ ({\it e.g.,} fixing one of the $\alpha_i$ or one of the $\beta_t$ equal to zero) to avoid perfect collinearity of the regressors, but this normalization does not affect the value for the estimator of the parameter of interest,  $\tau.$

Under a block assignment structure we have  $W_{it}=1$ only for a subset of the units (the ``treatment group'' with $i\in{\cal I}$, where the cardinality for the set ${\cal I}$ is $N^\treat$, and $N^\control\equiv N-N^\treat$), and  those units are treated  only during periods $t$ with $t>T_0$ (``post-treatment'').
Defining the averages in the four groups as
\[ \oy^{\treat,\post}\equiv \frac{\sum_{i\in{\cal I}}\sum_{t>T_0} Y_{it}}{N^\treat (T-T_0)},
\quad  \oy^{\treat,\pre}\equiv \frac{\sum_{i\in{\cal I}}\sum_{t\leq T_0} Y_{it}}{N^\treat T_0},
\]
\[ \oy^{\control,\post}\equiv \frac{\sum_{i\notin{\cal I}}\sum_{t>T_0} Y_{it}}{N^\control (T-T_0)},
\quad\textrm{and } \quad \oy^{\control,\pre}\equiv \frac{\sum_{i\notin{\cal I}}\sum_{t\leq T_0} Y_{it}}{N^\control T_0},
\]
 we can write the estimator for the treatment effect as
\[ \hat\tau^\twfe= \hat\tau^\did=\Bigl(\oy^{\treat,\post}- \oy^{\treat,\pre}\Bigr)-\Bigl(\oy^{\control,\post}-\oy^{\control,\pre}\Bigr),\]
in the familiar double difference form that motivated the DID terminology.

It is convenient to use the TWFE characterization based on least squares estimation of the regression function in (\ref{eq:twfe_obs}) because this characterization also applies in settings where the estimator does not have the double difference form, including the staggered adoption setting and even more general assignment processes. For this reason we also generally use the TWFE rather than the DID terminology in the remainder of this discussion.

\subsection{The Difference-In-Differences Estimator in the Grouped Repeated Cross-Section Setting}

Here we study the grouped repeated cross-section case where we observe each physical unit only once, obviously implying that we observe different units at different points in time. We continue to focus on the case with  the blocked assignment. To reflect this, our notation now only has a single index for the unit, $i=1,\ldots, N$. Let $G_i\in{\cal G}=\{1,\ldots,G\}$ denote the cluster or group unit $i$ belongs to, and $T_i\in\{1,\ldots,T\}$ the time period unit $i$ is observed in. The set of clusters ${\cal G}$ is partitioned into two groups, a control group ${\cal G}_C$ and a treatment group ${\cal G}_T$, with cardinality $G_C$ and $G_T$ respectively.

Units belonging to a group $G_i$ with $G_i \in{\cal  G}_C$  are not exposed to the treatment, irrespective of the time the units are observed. Units with $G_i \in{\cal G}_{T}$ are exposed to the treatment if and only if they are observed after the treatment date $T_0$, so that the treatment indicator is $W_{i}=\mathbf{1}_{G_i\in{\cal G}_T,T_i>T_0}$. Let $D_i=\mathbf{1}_{G_i\in{\cal G}_T}$ be the treatment group indicator that indicates whether unit $i$ is in one of the treated groups, irrespective of whether this unit is observed in the post-treatment period, so that $W_i=D_i\mathbf{1}_{T_i>T_0}$.
 
To define the DID estimator 
we first average outcomes and treatments for all units within a cluster/time period and construct $\overline{Y}_{gt}$ and $\overline{W}_{gt}$.  By assumption that the treatment within group and time period pairs is constant, the cluster/time-period average treatment $\overline{W}_{gt}$ is binary if the original treatment is.
The DID estimator is then the double difference
\[ \hat\tau^\did=\frac{1}{G_T (T-T_0)}
\sum_{g\in {\cal G}_T,t>T_0} \oy_{gt}
-\frac{1}{G_C (T-T_0)}
\sum_{g\in {\cal G}_C,t>T_0} \oy_{gt}\]
\[
-\frac{1}{G_TT_0}
\sum_{g\in {\cal G}_T,t\leq T_0} \oy_{gt}
+\frac{1}{G_C T_0}
\sum_{g\in {\cal G}_C,t\leq T_0} \oy_{gt}
\]
Alternatively we can use the TWFE specification at the group level (it cannot be used at the unit level because we do not observe any unit multiple times). At the group level we do have a proper panel setup:
\begin{equation}\label{gwfe2} \overline{Y}_{gt}(0)=\alpha_g+\beta_t+\varepsilon_{gt},\hskip1cm \overline{Y}_{gt}(1)=\overline{Y}_{gt}(0)+\tau,\end{equation}
similar to that in (\ref{eq:twfe}). The potential outcomes $\overline{Y}_{gt}(0)$ and $\overline{Y}_{gt}(1)$ should here be interpreted as the average of the potential outcomes if all units in a group/time-period pair are exposed to the control (active) treatment. The group-level TWFE estimator is identical to the DID estimator.

\subsection{Inference}\label{section:did_inference}

To conduct inference about $\hat \tau^\did$ or $\htautwfe$ we need to be explicit about the sampling and assignment schemes. In situations where the assignment process is known, such as in randomized experiments, we can do design-based or randomization-based inference. We discuss this approach in detail in Section \ref{section:design}. Outside of such situations, researchers typically rely on sampling-based inference, which we outline below.

In the proper panel setting, one often assumes that all units are randomly sampled from a large population and thus exchangeable. In this case, inference about $\htautwfe$ reduces to joint inference about four means with i.i.d. observations. This approach was used by \citep{card1994minimum} to quantify the uncertainty about the estimated effect of the minimum wage.

With GRCS data and the cardinality of  the
control and treatment groups ${\cal G}_C$ and ${\cal G}_T$ larger than one, the situation is different. Now in addition to accounting for variation within a group at the unit level, one can allow for non-vanishing errors at the group level, the $\varepsilon_{gt}$ in the model in Equation (\ref{gwfe2}). 
This cannot be done in the two-group/two-period case as in \cite{card1994minimum} because one cannot estimate the between-group variation in the presence of group fixed effects, and mechanically the estimated residuals $\hat\varepsilon_{gt}$ are all equal to zero.
The clustering approach allowing for non-vanishing $\varepsilon_{gt}$ was advocated in  \citep{liang1986longitudinal, manuel1987computing, Bertrand2004did, donald2007inference, ibragimov2016inference}, and is routinely used in situations where the number of groups or periods exceeds two. See \citep{abadie2023should} for a recent discussion in a design setting.

In addition to accounting for the presence of non-vanishing $\varepsilon_{gt}$, since  \citep{Bertrand2004did} inference for TWFE estimators has typically taken into account the correlation in outcomes over time within units in applications with more than two periods. This implies that if one estimates the average treatment effect as in (\ref{estimator:twfe}), it is not appropriate to use the robust, Eicker-Huber-White standard errors. Instead, one can use clustered standard errors (\citealp{liang1986longitudinal, manuel1987computing}), based on clustering observations by units.
The appropriate standard errors can also be approximated by bootstrapping all observations for each unit. \cite{hansen2007generalized} discusses a more general hierarchical setup.

\subsection{The Parallel Trend Assumption}\label{sec:par_trends}

The fundamental justification for the TWFE estimator, in one form or another, is based on a parallel trend assumption. This states that, in one form or another, the units who are treated would have followed, in the absence of the treatment, a path that is parallel to the path followed by the control units, in an average sense. The substantive content and the exact form of the assumption depend on the specific setup, the proper panel case versus the grouped repeated cross-section case, whether one takes a model-based or design-based perspective, the number of groups, and the averaging that is performed.

Let us first consider the proper panel case, with block assignment and  $D_i$ the indicator for the event that unit $i$ will be exposed to the treatment in the post-treatment period (after $T_0$).
Then, the assumption is that the expected  difference  in control outcomes in any period for  units who later are exposed to the treatment and units who are always in the control group is constant:
\begin{assumption}
For all  $t,t'$,
\begin{equation}
\mme[Y_{it}(0)|D_i=1]-\mme[Y_{it}(0)|D_i=0] = \mme[Y_{it'}(0)|D_i=1]-\mme[Y_{it'}(0)|D_i=0] . \end{equation}
\end{assumption}
Equivalently we can formulate this assumption in terms of changes over time. In that formulation, the assumption is that the expected change in control outcomes is the same for those who will eventually be exposed to the treatment and those who will not:
\[
\mme [Y_{it}(0)-Y_{it'}(0)|D_i=1] = \mme [Y_{it}(0)-Y_{it'}(0)|D_i=0] \hskip1cm \forall t,t'. \]
To motivate this assumption for the panel case an alternative is to postulate a TWFE model for the control outcomes, as in (\ref{eq:twfe_obs}),
with the additional assumption that the treatment assignment $D_{i}$ is independent of the vector of residuals $\varepsilon_{it}$, $t=1,\ldots,T$ conditional on fixed effects:
\[ D_{i}\ \indep\ (\varepsilon_{i1},\ldots,\varepsilon_{iT}) | \alpha_i,\]
as in, for example,  \citep{arellano2003panel}.
From the point of view of the modern causal inference literature, the parallel trend assumption is somewhat non-standard because it combines restrictions on the potential outcomes with restrictions on the assignment mechanism (see \cite{ghanem2022selection, roth2023parallel} for  additional discussion).

Consider next the Grouped Repeated Cross-Section (GRCS) case. Suppose in the population all groups are large (infinitely large) in each period, and we have random samples from these populations for each period. Then the expectations are well defined as population averages.  In that case, the parallel trends assumption can be formulated as requiring that the difference in expected control outcomes between two groups remains constant over time:
\begin{assumption}
For all pairs of groups $g,g'$ and for all pairs of time periods $t,t'$, the average difference between the groups remains the same over time, irrespective of their treatment status:
\begin{equation}
\mme\Bigl[Y_{gt}(0)\Bigl| D_i=1\Bigr] - \mme\Bigl[Y_{g't}(0)\Bigl| D_i=0\Bigr] = \mme\Bigl[Y_{gt'}(0)\Bigl| D_i=1\Bigr] - \mme\Bigl[Y_{g't^{\prime}}(0)\Bigl| D_i=0\Bigr]. 
\end{equation}
\end{assumption}
Again an alternative formulation is as the assumption that the expected change between periods $t'$ and $t$ is the same for all groups:
\[
\mme\Bigl[Y_{gt}(0)\Bigl| D_i=1\Bigr] -  \mme\Bigl[Y_{gt^{\prime}}(0)\Bigl| D_i=1\Bigr] = \mme\Bigl[Y_{g't}(0)\Bigl| D_i=0\Bigr] -\mme\Bigl[Y_{g't'}(0)\Bigl| D_i=0\Bigr],
\]
 for all $ g,g',t,t'$.
If we were to observe $Y_{gt}(0)$ for all groups and time periods, then the presence of two groups and two time periods would be sufficient for this assumption to have testable implications. However, with at least one of the four cells defined by the group and time period exposed to the treatment, there are no testable restrictions implied by this assumption in the two-group / two-period case, as in for example in the New-Jersey/Pennsylvania minimum wage study in \citep{card1994minimum}.

Because we can view the panel case as a two-group setting, with the defined in terms of the indicator $D_i\in\{0,1\}$, there are only testable restrictions from this assumption when we have more than two periods. With more than two groups, just as in the case with more than two periods, there are testable restrictions implied by the parallel trend assumption. In an early paper \citep*{ashenfelter1985using} argued against using the TWFE model for evaluation of training programs based on the failure of parallel trends detected in the data. See \citep*{roth2023s,rambachan2023more, jakiela2021simple} for a discussion and bounds based on limits on the deviations from parallel trends.
Bridging some of the gap between design and sampling-based approaches
 \citep{roth2023parallel} show how parallel trends can be implied by random assignment of treatment. They also discuss the sensitivity to transformations of the parallel trend assumption. We return to this in Section \ref{section:nonlinear}, where we discuss nonlinear methods.

\subsection{Pre-treatment Variables}

Often researchers observe time-invariant characteristics of the units in addition to the time path of the outcome. Such characteristics cannot be incorporated simply by adding them to the TWFE specification in (\ref{eq:twfe_obs}) because they would be perfectly colinear with the individual components $\alpha_i$. Nevertheless, the pre-treatment variables can be important by 
facilitating a decomposition of these effects into explained components as in 
 (\citep{plumper2007efficient}), or by allowing the relaxation of some of the key assumptions. Specifically, one can assume that the parallel trend and constant treatment effect assumptions hold only within subpopulations defined by these characteristics. Two specific proposals have been made for the applications with time-invariant covariates. 

\subsubsection{Abadie (2005)}

In an early paper \citep{abadie2005semiparametric} proposes flexible ways of adjusting for time-invariant covariates while continuing with a conditional version of the parallel trends assumption. His solution was based on re-weighting the differences in outcomes by the propensity score to ensure balance.

\subsubsection{Sant’Anna and Zhao (2020)}

\citep{sant2020doubly} use recent advances in the cross-section causal inference literature to adjust for time-invariant covariates in a doubly robust way by combining inverse-propensity score weighting with outcome modeling. In cross-sectional settings, such doubly robust methods have been found to be more attractive than either outcome modeling or inverse-propensity-score weighting on their own. They do maintain the parallel trends assumption conditional on covariates.\\

With finite $T$, strictly exogenous time-varying covariates $X_{it}$ can be converted to time-invariant $X_{i} \equiv (X_{i1},\dots, X_{iT})$. Applied researchers rarely follow this practice and instead rely on linear specifications with contemporaneous covariates. For a discussion of the problems with the conventional specifications and a potential solution, see \citep{caetano2022difference}.

\subsection{Unconfoundedness}\label{section_unconf}

One key distinction between the repeated cross-section and proper panel case (and also the grouped-repeated-cross-section case after aggregration) is that 
in the case with proper panel data there is a natural alternative to the TWFE estimator. This is most easily illustrated in the case with blocked assignment, where the treatment group is only exposed to the treatment in the last period. Viewing the pre-treatment outcomes as covariates, one could assume unconfoundedness:
\begin{equation}
 D_i\ \indep\ \Bigl(Y_{iT}(0),Y_{iT}(1)\Bigr)\ \Bigl|\ 
 Y_{i1},\ldots,Y_{iT-1}.
\end{equation}
If one is willing to make this assumption, the larger literature on the estimation of treatment effects under unconfoundedness applies. See
\citep{imbens2004} for a survey. Modern methods include doubly robust methods that combine modeling outcomes with propensity score weighting. 
See, for example, \citep{bang2005doubly,  chernozhukov2017double, athey2018approximate}.

Consider the case with two periods, $T=2$.
Because unconfoundedness is equivalent to assuming
\[D_i\ \indep\ \Bigl(Y_{i2}(0)-Y_{i1},Y_{i2}(1)-Y_{i1}\Bigr)\ \Bigl|\ 
 Y_{i1},\]
 it follows that the issue in the choice between TWFE and unconfoundedness is really whether one should adjust for differences 
 between treated and control units in the lagged outcome, $Y_{i1}$. The TWFE approach implies one should not and that doing so may introduce biases that are otherwise absent, and the unconfoundedness approach implies one should adjust for these differences. 

The unconfoundedness assumption and TWFE model validate different non-nested comparisons and applied researchers often do not carefully and explicitly motivate their choices in this regard. The key difference between the two models is the underlying selection mechanism. The TWFE model assumes that the treated units differ from the control ones in unobserved characteristics that are potentially correlated with a persistent component of the outcomes -- the fixed effect $\alpha_i$. The unconfoundedness assumption, on the other hand, is satisfied when the selection is based solely on past rather than future outcomes (and potentially other observed pre-treatment variables). 
 
 The methodological literature does not provide a lot of guidance on the choice between these two strategies, with exceptions in \citep{angristpischke,  xu2023causal}. It is somewhat segmented, with some subliteratures focusing solely on fixed effect strategies and some solely focusing on unconfoundedness approaches. For example, there is a large literature re-analyzing the data originally studied in
\citep{lalonde1986evaluating} (see also \citep{dehejiawahba, imbens2024lalonde}) where the researcher has observations on two lagged outcomes. Although LaLonde reports estimates from various TWFE models in addition to estimates that adjust for initial period outcomes, in the subsequent literature the focus is almost entirely on methods assuming unconfoundedness. 
In contrast, most of the literature reanalyzing the data originally studied in \citep{card1994minimum} where the researcher observes outcomes for a single pre-treatment period has focused on TWFE and related methods with relatively little attention paid to unconfoundedness approaches. It is not clear what motivates the differences in emphasis in these two applications.
In an early study, \citep{ashenfelter1985using} carefully point out the limitations of the TWFE model and, in particular, its inability to capture temporary declines in earnings prior to enrollment in labor market programs, the so-called Ashenfelter dip. In our view the unconfoundedness approach is perhaps under-utilized in the empirical panel data literature with the case for the fixed effect specification overstated.

There are two important cases where the unconfoundedness and TWFE approaches lead to similar results. Again, this is most easily seen in the two-period case. The results from the two approaches are similar
if the averages of the initial period outcomes are similar for the two groups {\it or} if the average in the control group did not change much over time.
One way to think about this case is to view it as one where there are multiple potential control groups. One can use the contemporaneous control group, or one can use the treatment group in the first period. If either the control group does not change over time, or if the treatment group and the control group do not differ in the first period, then the two potential control groups deliver the same results.
See for more on this multiple control group perspective \citep{rosenbaum2002multiple}.

When the control group changes over time, and in addition the control group and treatment group differ in the initial period, then the TWFE and unconfoundedness approaches give different results. However, the differences can be bounded, albeit under additional assumptions.  Suppose unconfoundedness holds and the distribution of the pretreatment outcomes in the treatment group stochastically dominates that in the control group. Then, the TWFE estimator will underestimate the true effect. On the other hand, if the TWFE model holds, then assuming unconfoundedness and adjusting for the lagged outcome will overestimate the true effect.  See Chapter 5.4 in \cite{angristpischke} for a derivation in the linear case and 
\cite{ding2019bracketing} for a nonparametric generalization that allows for heterogeneity in treatment effects, {\it i.e.}, failure of Assumption \ref{ass:constant}. 
\citep{imai2021matching} takes a middle ground between unconfoundedness assumptions and TWFE/DID methods by conditioning on lagged outcomes other than the most recent one, which is differenced out in a TWFE approach.

\subsection{Distributional effects}
Our discussion in this section has focused on an average effect $\tau$.  This is without essential loss of generality as long as Assumption \ref{ass:constant} holds, but in practice, researchers do not expect this assumption to hold exactly. As we discuss in the next section, heterogeneity in treatment effects does not create problems for the DiD estimator, which continues to estimate an interpretable average treatment effect. At the same time, when treatment effects are heterogeneous, researchers can be interested in estimands that capture the distributional effects of the treatment, and panel data provides the opportunity to estimate such effects.

In particular, in \cite{bonhomme2011recovering}, the authors show how to use deconvolution techniques to identify the full distribution of the treatment effects for the treated units, as long as the treatment effects and treatment assignment $W_{it}$ are statistically independent of the idiosyncratic errors $\varepsilon_{it}$ introduced in Assumption \ref{ass:twfe}. Note that this restriction does not put any structure on the correlation between the assignment, unit fixed effects $\alpha_i$, and treatment effects, thus allowing for rich selection patterns, {\it e.g.}, selection on the treatment effects themselves. 

In \cite*{callaway2018quantile,callaway2019quantile}, the authors use a different strategy and show that as long as the difference in outcomes is not correlated with the treatment assignment, researchers can identify quantile treatment effects under additional stability restrictions on the joint distribution of the potential outcomes. Their approach is connected to the Changes-In-Changes setup in \citep{athey2006identification}, which we discuss in more detail in Section \ref{section:nonlinear}.

\section{The Staggered Adoption Case}\label{section:staggered}

Although much of the panel literature starts with the TWFE model for control outcomes (Assumption \ref{ass:twfe}) with constant treatments effects (Assumption \ref{ass:constant}), the constant treatment effect assumption is not important in the setting with block assignment. Maintaining the TWFE for control outcomes but allowing unrestricted heterogeneity in the treatment effects $Y_{it}(1)-Y_{it}(0)$, the TWFE estimator (which then has the double difference form) continues to estimate a well-defined average causal effect, namely the average treatment effect for the treated in the periods in which they received the treatment,
\[ \frac{1}{\sum_{i=1}^N\sum_{t=1}^T W_{it}}\sum_{i=1}^N \sum_{t=1}^T W_{it}\Bigl( Y_{it}(1)-Y_{it}(0)\Bigr).\]
The interpretation is more complex in settings with dynamic treatment effects, but the underlying estimand is still well-defined. 
This robustness to treatment effect heterogeneity does not extend to settings outside of block assignment.

Part of the new causal panel literature 
builds on traditional TWFE methods in the staggered adoption setting, allowing for general heterogeneity in the treatment effects.
The twin goals are understanding what is estimated by TFWE estimators in this setting that is common in empirical work (see \citealp{de2023two} for evidence on how common this setting is), and proposing modifications that ensure that a meaningful average causal effect is estimated.
We review that literature here.
Recall the staggered adoption setting,
\[
\begin{array}{r}{\bw^\stag=} \\
{(\text{staggered  adoption})}
\end{array}
\left(
\begin{array}{ccccccccc}
	\ttick &  \ttick & \ttick  & \ttick & \dots & \ttick & \ttick \\
	\ttick  & \ttick &  \ttick & \ttick   & \dots  & \ttick& {\ttock}  \\
	\ttick  & \ttick &  \ttick  & \ttick  & \dots  & \ttock& {\ttock}  \\
	\ttick  & \ttick & \ttick &  {\ttock}   & \dots & \ttock & {\ttock}  \\
	\vdots   &  \vdots  &  \vdots& \vdots &\ddots &\vdots &\vdots \\
	\ttick  & {\ttock} &  {\ttock}   & {\ttock} & \dots  & \ttock& {\ttock}  \\
\end{array}
\right)
\quad\ \quad
\begin{array}{c}\textrm{(never adopter)}
\\ \textrm{(very late adopter)}
\\ \textrm{(late adopter)}
\\ \textrm{(medium adopter)}
\\ \vdots
\\ \textrm{(early adopter)}
\end{array}
\]
Let $A_i\equiv T+1-\sum_{t=1}^T W_{it}$ be the adoption date (the first time unit $i$ is treated if a unit is ever treated), with the convention that $A_i\equiv \infty$ for units who never adopt the treatment, and recall that $N_a$ is the number of units with adoption date $A_i=a$.
Define also the average treatment effect by time and adoption date,
\[ \tau_{t|a}\equiv \mme\left[ \left.Y_{it}(1)-Y_{it}(0)\right|A_i=a\right].\]
The key is that these average treatment effects can vary both by time and by adoption date.
Such heterogeneity was rarely allowed for in the earlier literature, with an early exception in \citep{chamberlain1992efficiency} and more recently in \citep{arellano2011identifying,graham2012identification,chernozhukov2013average}. We discuss the connection between this literature and the modern one at the end of this section. 
Note that in this setting, we cannot separate the presence of dynamic effects from heterogeneity in the treatment effects over time and by adoption date.

\subsection{Decompositions of the TWFE Estimator}

Here we discuss the interpretation of the TWFE estimator $\hat\tau$ based on the least squares regression
\[ \min_{\alpha,\beta,\tau}\sum_{i=1}^N \sum_{t=1}^T \left(Y_{it}-\alpha_i-\beta_t-\tau W_{it}\right)^2,\]
in the staggered adoption case. This decomposition is based on 
the discussion in  \citep{goodman2021difference}. We maintain Assumption \ref{ass:twfe}, which implies the TWFE structure for the control outcomes and the mean-independence between the residuals and the treatment indicator.

Define for all time-periods $t$ and all adoption dates $a$ the average outcome in period $t$ for units with adoption date $a$:
\[ \oy_{t|a}\equiv \frac{1}{N_a}\sum_{i:A_i=a} Y_{i,t}.\]
Then, for all pairs of time periods $t>t'$ and pairs of adoption dates $a,a'$ such that $t'<a\leq t$ (units with adoption date $a$ change treatment between $t$ and $t'$) and either $a'\leq t'$ or $t<a'$ (units with adoption date $a'$ do not change treatment status between $t$ and $t'$, they are either already treated before period $t'$, or only adopt the treatment after period $t$), define the following double difference that is the building block for the  TWFE estimator:
\begin{equation}\label{eq:gb} \hat\tau_{t,t'}^{a,a'}\equiv \Bigl(\overline{Y}_{t|a}-\overline{Y}_{t^{\prime}|a}\Bigr)
-\Bigl(\overline{Y}_{t|a'}-\overline{Y}_{t'|a'}\Bigr)\end{equation}
The interpretation of this double difference plays a key role in the interpretation of the TWFE estimator $\hat\tau$. The group with adoption date $a$ changes treatment status between periods $t'$ and $t$, so the difference 
$\overline{Y}_{t|a}-\overline{Y}_{t'|a}$ reflects a change in treatment but this treatment effect is contaminated by the time trend in the control outcome under the TWFE structure:
\[\mme\left[\overline{Y}_{t|a}-\overline{Y}_{t'|a'}\right]=\beta_{t}-\beta_{t'}+\tau_{t|a}.\]For the group with an adoption date $a'$, the difference
$\overline{Y}_{t|a'}-\overline{Y}_{t'|a'}$ does not capture a change in treatment status. If $t<a'$, it is a difference in average control outcomes, and $\hat\tau_{t,t'}^{a,a'}$ is a standard DID estimand, which under the TWFE model for the control outcomes has an interpretation as an average treatment effect. 
\citep*{roth2023s} refer to this as a ``clean'' comparison.

However, if $a'<t'$,  the difference
$\overline{Y}_{t|a'}-\overline{Y}_{t'|a'}$ is a difference in average outcomes under the treatment. In the presence of treatment effect heterogeneity, and in the absence of a TWFE model for the outcomes under treatment, its expectation can be written as
\[\mme\left[\overline{Y}_{t|a'}-\overline{Y}_{t'|a'}\right]=\beta_{t}-\beta_{t'}+\Bigl(\tau_{t|a'}-\tau_{t'|a'}\Bigr).\]
Hence, in the case with $a'<t'$, the basic building block in (\ref{eq:gb}) has expectation
\[\mme\left[\hat\tau_{t,t'}^{a,a'}\right]=\tau_{t|a}-\left(\tau_{t|a'}-\tau_{t'|a'}\right).\]
This is a weighted average of treatment effects, with the weights adding up to one but with some of the weights negative. This is sometimes referred to as a ``forbidden'' comparison (\cite{roth2023s}).
If the treatment effects are all identical, this does not, in fact, create a concern. However,  if there is reason to believe there is substantial heterogeneity, as is likely in practice, researchers may be reluctant to report weighted averages with negative weights.
Note that the concern with the comparisons $\hat\tau_{t,t'}^{a,a'}$ when $a'<t'$ but not when $a'>t$ fundamentally treats the treated state and the control state asymmetrically: the parallel trends assumption is maintained for the control outcomes, but not for the treated outcomes.

The TWFE estimator $\htautwfe$  can be characterized as a linear combination of the building blocks $\hat\tau_{t,t'}^{a,a'}$, including those where the non-changing group has an early adoption date $a'<t'$. The coefficients in that linear combination depend on various aspects of the data, including the number of units $N_a$ in each of the corresponding adoption groups, as discussed in detail in \citep{goodman2021difference,
baker2021much} for the staggered adoption case and in \citep*{imai2021use} for the general case. As a result, the TWFE estimator has two distinct problems. First, without further assumptions, the estimator does not have an interpretation as an estimate of an average treatment effect with nonnegative weights in general. Second, the combination of weights on the building blocks chosen by the TWFE regression depends on the data, in particular on the distribution of units across the adoption groups. As a result, two identical populations in terms of potential outcome distributions (and thus identical treatment effect distributions) that have different adoption patterns would lead to different estimated quantities. 

We emphasize that the expectations above are computed with respect to the errors $\varepsilon_{it}$ holding the adoption dates fixed. This is in line with the fixed-effects tradition in the panel data literature, which does not restrict the conditional distribution of unit-specific parameters, such as $\tau_{it}$, given the covariates of interest, which in our case corresponds to $A_{i}$. In some situations, {\it e.g.,} in randomized experiments, the adoption date is unrelated to the $\tau_{it}$ and thus the conditional distribution of the $\tau_{it}$ is equal to its marginal distribution,
 and the negative weights issue does not necessarily arise, {\it e.g.}, \citep{lihua2021}. We return to this point and its connection to the random-effects tradition in the panel data literature in Section \ref{section:design}.

We also note that in other settings, including linear regression, researchers often report estimates that in the presence of treatment effect heterogeneity represent weighted averages of treatment effects with some of the weights negative. While that is not necessarily ideal, there are in the current setup tradeoffs with other assumptions, including the parallel trend assumptions, that may force the researcher to make some assumptions that are, at best, approximations. Similar tradeoffs motivate the use of higher-order kernels in nonparametric regression, which also lead to estimators with negative weights. We, therefore, do not view the negative weights of some estimators as necessarily disqualifying. We also find the terminology ``clean'' and ``forbidden'' not doing justice to the potential benefits from such methods. 

\subsection{Alternative DID-type Estimators for the Staggered Adoption Setting}

To deal with the negative weights, researchers have recently, more or less contemporaneously, proposed a number of different modifications to the TWFE estimator. Here we discuss four of these modifications that have attracted considerable attention. It should be noted that all maintain the TWFE assumption for the control outcomes, and all four avoid the additional assumption on treatment effect heterogeneity.

\subsubsection{Callaway and Sant'Anna (2020)}

\citep{callaway2020difference}
propose two ways of dealing with the negative weights. Their first approach takes a group with adoption date $a$, and compares average outcomes in any post-adoption period $t\geq a $  ($\overline{Y}_{t|a}$ for $t\geq a$) to average outcomes for the same group (the group with adoption date $a$) immediately prior to the adoption ($\overline{Y}_{a-1|a}$). It then subtracts the difference in outcomes for the same two time periods for the single group that never adopts the treatment ($a = \infty$).
Formally, consider, for $t\geq a$, the double difference
\begin{equation}\label{calla}\hat\tau_{t,a-1}^{a,\infty}= \Bigl(\overline{Y}_{t|a}-\overline{Y}_{a-1|a}\Bigr)
-\Bigl(\overline{Y}_{t|\infty}-\overline{Y}_{a-1|\infty}\Bigr).\end{equation}
A concern is that this particular control group, those who never adopt the treatment, may not be particularly attractive. One might worry that the very fact that this group never adopts the treatment is an indication that they are fundamentally different from the other groups and thus less suitable as a comparison for the trends in the absence of the treatment.
In addition, very few of these never-adopters may exist, especially in long panels, so the precision of the estimators based on such comparisons may make them unattractive.

Recognizing this concern \citep{callaway2020difference} suggest 
using as an alternative control group the average of the groups that do adopt the treatment, but restricting this to those who adopt after period $t$:
\[\hat\tau_{t,a-1}^{a,>t}\equiv \Bigl(\overline{Y}_{t|a}-\overline{Y}_{a-1|a}\Bigr)
-\frac{1}{T-t}\sum_{a'=t+1}^T\Bigl(\overline{Y}_{t|a'}-\overline{Y}_{a-1|a'}\Bigr)\]

Given these two estimators, \citep{callaway2020difference} suggest reporting averages over periods $t$ and adoption dates $a$, using a variety of possible weight functions $\omega(a,t)$ that depend on the adoption date and the time period.  One of their preferred weight functions is
\[ \omega_e(a,t)=\mathbf{1}_{a+e=t}\cdot
\pr(A_i=a|A_i\leq T-e),
\]
which leads to an average of treatment effects, over different adoption dates, at exactly $e$ periods after adoption, for their two control groups,
\[ \hat\tau^{\textrm{CS,I}}(e)=\sum_{a=2}^{T-e} \omega_e(a,t)\cdot \hat\tau_{t,a-1}^{a,\infty},
\quad\textrm{
or}\quad
 \hat\tau^{\textrm{CS,II}}(e)=\sum_{a=2}^{T-e} \omega_e(a,t)\cdot \hat\tau_{t,a-1}^{a,>t}.
\]

We should note that
\citep{callaway2020difference} also allow for the possibility that the treatment is anticipated, and so that up to some known number of periods prior to the treatment, the outcome may already be affected by this.

\subsubsection{Sun and Abraham (2020)}

\citep{sun2020estimating} start with one of the same
building blocks as \citep{callaway2020difference}, $\hat\tau_{t,a-1}^{a,\infty}$ in
(\ref{calla}).
Given double differences of this type they suggest reporting the average of this:
\[ \hat\tau^{\textrm{SA}}=\sum_{t=2}^T\sum_{a=2}^t\hat\tau_{t,a-1}^{a,\infty}\cdot \frac{\pr(A_i=a|2\leq A_i\leq t)\cdot 
\mathbf{1}_{2\leq a\leq t\leq T}}{T-1}.\]
This is a simple unweighted average over the periods $t$ after the first period, with the weights within a period equal to the fraction of units with an adoption date prior to that, excluding first period adopters.

An  additional issue emphasized by \citep{sun2020estimating} is related to the validation of the two-way model. In applications, this validation is done by testing for parallel trends using pre-treatment data. \citep{sun2020estimating} show that common implementation of such tests using two-way specifications with leads of treatments also include  comparisons with negative weights. As a result, they caution against such procedures.

\subsubsection{de Chaisemartin and d'Haultf{\oe}uille, 2020}

\citep{de2020two}
deal with the negative weights by focusing on one-period ahead double differences, with control groups that adopt later ($a>t$):
\[\hat\tau_{t,t-1}^{t,a}=\Bigl(\overline{Y}_{t|t}-\overline{Y}_{t-1|t}\Bigr)
-\Bigl(\overline{Y}_{t|a}-\overline{Y}_{t-1|a}\Bigr).\]
They aggregate these by averaging over all groups that adopt later:
\[\hat\tau_{+,t}=\frac{1}{T-(a-1)}\sum_{a>t} \hat\tau_{t,t-1}^{t,a}.\]
Then they average over the time periods, weighted by the fraction of adopters in each period:
\[ \hat\tau^{\textrm{CH}}=\sum_{t=2}^T\hat\tau_{+,t}\cdot\pr(A_i=a|A_i\geq 2).\]
One challenge with the \citep{de2020two} approach is that by limiting the comparisons to those that are separated by a single period, the standard errors may be large relative to those for estimators based on more comparisons. Although the additivity assumption may be more likely to hold over such short horizons, there is also increased sensitivity to the presence of dynamic effects.

\subsubsection{Borusyak, Jaravel, and Spiess, 2021}

 \citep{borusyak2021revisiting} focus on a  model for the baseline outcomes that is richer than the TWFE model:
 \begin{equation*}
     Y_{it}(0) = A_{it}^\top \lambda_i + X_{it}^\top \delta + \epsilon_{it}
 \end{equation*}
 where $A_{it}$ and $X_{it}$ are observed covariates, leading to a factor-type structure. This setup reduced to the TWFE for $A_{it} \equiv 1$ and $X_{it} \equiv \left(\mathbf{1}_{t = 1}, \dots, \mathbf{1}_{t = T}\right)$. They propose estimating $\lambda_i$ and $\delta$ by least squares using only observations for control units only, and later construct unit-time specific imputations for the unobserved control outcomes for the treated units, leading to unit/period-specific treatment effect estimates:
 \begin{equation*}
     \hat \tau_{it} = Y_{it} -  A_{it}^\top \hat\lambda_i + X_{it}^\top \hat\delta.
 \end{equation*}
These unit-specific estimators can then be aggregated into an estimator for the target of interest; let us call the estimator $\hat\tau^\mathrm{BJS}$. Notably, despite each
unit-time specific treatment effect estimator $\hat \tau_{it}$ being inconsistent, after these objects are averaged, the estimator is well-behaved. Moreover,  \citep{borusyak2021revisiting} show that the resulting estimator is efficient as long as $\epsilon_{it}$ is i.i.d. over $i$ and $t$, which relies on a version of the Gauss-Markov theorem for their setup.

\subsubsection{Discussion}

If one is concerned with the negative weights in the TWFE estimator in a setting with staggered adoption, how should one choose between these four alternatives, 
$\hat\tau^{\textrm{CS,I}}$ (or $\hat\tau^{\textrm{CS,II}}$), $\hat\tau^{\textrm{SA}}$, 
$\hat\tau^{\textrm{CH}}$, and $\hat\tau^\mathrm{BJS}$? The first key issue is the choice of the estimand. In staggered designs there are many average effects one can estimate, and the choice of which one to report should be addressed carefully depending on the underlying research question. Once this choice is made, there are some substantive arguments that matter for the choice of the estimator: \begin{inparadesc}\item[$(i)$] the never-adopter group may well be substantively different from groups that eventually adopt, \item[$(ii)$] for long differences (where we compare outcomes for time periods far apart) the assumption that differences between units are additive and stable over time becomes increasingly less plausible, \item[$(iii)$] one-period differences may be quite different from differences based on comparisons separated by multiple periods if there are dynamic effects, and \item[$(iv)$] efficiency considerations.
\end{inparadesc}
These concerns do not lead to one proposal clearly dominating the others, and in practice, looking for a single estimator may be the wrong goal.

What should one do instead? One option is to report all of the proposed estimators, as, for example, \citep{braghieri2022social}, who report estimates based on all four approaches in addition to the standard TWFE estimator.
However, that does not do justice to the fact that the estimators rely on fundamentally different assumptions, in particular about treatment effect heterogeneity, and focus on different estimands. Moreover, some of these comparisons may have little power in terms of uncovering heterogeneity of particular forms. Finally, other than \citep{borusyak2021revisiting}, the methods all rely on some version of parallel trend assumptions. Ultimately, instead of reporting all estimators, we therefore recommend exploring directly the presence of systematic variation in the $\hat\tau_{t,t'}^{a,a'}$, by adoption date, $a$, by the length of the period between before and after, $t-t'$, and the time since adoption, $t-a$.  

\subsubsection{Relation to earlier literature}

Here we relate the discussion in this section to some earlier results in econometric panel data literature. In \citep{chamberlain1992efficiency}, \citep{graham2012identification} and \citep{arellano2011identifying}, the authors analyze a class of panel data models that incorporates the two-way model with heterogeneous treatment effects as a special case. For example, \citep{arellano2011identifying} postulate the following model:
\begin{equation}\label{eq:ab_mod}
    \mathbf{Y_i} = \mathbf{Z}_i \delta + \mathbf{X}_i \gamma_i + \boldsymbol{\varepsilon}_{i}, \quad \mathbb{E}[\boldsymbol{\varepsilon}_{i}|\gamma_i, \mathbf{Z}_i, \mathbf{X}_i] = 0,
\end{equation}
where $ \mathbf{Y_i}$ is a $T$-dimensional vector of outcomes, and $\mathbf{Z}_i$ and $\mathbf{X}_i$ are matrices of regressors for unit $i$ and $\boldsymbol{\varepsilon}_{i}$ is a $T$-dimensional vector of errors.  To see that this model includes those discussed in this section, first define $Z_i = \mathcal{I}_{T}$, a $T\times T$ identity matrix, and $\delta = (\beta_1, \dots, \beta_{T})^\top$. Next, define 
\begin{align*}
    \mathbf{X}_i =\begin{pmatrix} 1 & W_{i,1} & 0& \dots& 0 \\
                                  1 & 0 & W_{i,2} & \dots& 0\\
                                  \dots \\
                                  1 & 0 & 0 & \dots & W_{i,T}\end{pmatrix}
\end{align*}
and $\gamma_i = (\alpha_i, \tau_{i1},\dots, \tau_{iT})^\top$. Equation \eqref{eq:ab_mod} then reduces to the two-way model with heterogeneous treatment effects:
\begin{equation*}
    Y_{it} = \alpha_i + \beta_t + \tau_{it}W_{it} + \varepsilon_{it}.
\end{equation*}
A similar relation applies to the setup described in \citep{graham2012identification}. Both of these models are a particular instance of the setup described in Section 4 of \citep{chamberlain1992efficiency}.

The earlier econometric literature did not focus on the properties of the fixed effects estimator for a misspecified version of \eqref{eq:ab_mod} but instead was concerned with directly estimating distributional characteristics of $\gamma_i$, such as $\mathbb{E}[\gamma_i]$ in \citep{chamberlain1992efficiency,graham2012identification}, or $\mathbb{V}[\gamma_i]$ and higher-order moments in \citep{arellano2011identifying}. Because of this, the key assumption in these papers is the presence of units for which the matrix $\mathbf{X}_{i}^\top \mathbf{X}_i$ has a full rank. This assumption is needed to impute the value of $\gamma_i$:
\begin{align*}
    \hat \gamma_i = \left(\mathbf{X}_{i}^\top \mathbf{X}_i\right)^{-1}\mathbf{X}_i^\top(\mathbf{Y}_i - \mathbf{Z}_i \hat\delta),
\end{align*}
where $\hat \delta$ is a consistent estimator for the common parameter $\delta$, which is typically available. 

Because of the dimension of $\mathbf{X}_i$, this approach is infeasible in the two-way model with heterogeneous treatment effects, and it is impossible to identify any distributional characteristics of $\gamma_i$ for any subpopulation of units without additional assumptions. At the same time, often we are not interested in the distributional characteristics of the entire vector $\gamma_i$ but instead focus on components thereof, such as the average treatment effect in the periods when units are treated. For such estimands the results are  more positive as illustrated by the causal panel data literature. From this perspective, one can view the strategies discussed above, in particular, the imputation approach of \citep{borusyak2021revisiting}, as an extension of \citep{arellano2011identifying} to settings where only some components of $\gamma_i$ can be estimated.

\citep{chernozhukov2013average} is another example from the econometric panel data literature that emphasizes the heterogeneity in treatment effects. For cases with binary regressors, their model  has the following structure:\footnote{For binary regressors, this form is equivalent to the nonseparable model described in the paper. The assumptions the authors make imply additional distributional restrictions on $\varepsilon_{it}$, which are not needed for the estimation of average effects.}
\begin{equation*}
    Y_{it} = \beta_t + \lambda_t\left(\alpha_i + \tau_i W_{it}\right) + \varepsilon_{it}, \quad \mathbb{E}[\varepsilon_{it}|\alpha_i, \tau_i, W_{i1},\dots, W_{iT}] = 0.  
\end{equation*}
This model allows for more flexible baseline potential outcomes generalizing Assumption \ref{ass:twfe}. This aspect connects it to factor models we discuss in Section \ref{section:relax}. At the same time, the heterogeneity in treatment effects is limited and connected to heterogeneity in the baseline outcomes, unlike in other models discussed in this section.  The authors show that certain average treatment effects can be estimated in this model as long as there is enough variation in the treatment $W_{it}$. These results would typically apply to staggered designs as long as there are at least two different adoption periods. The authors also discuss the estimation of quantile effects, which relies on additional distributional restrictions on $\varepsilon_{it}$.

\subsection{Modeling heterogeneity}
In Section \ref{notation}, we discussed that the distinctive feature of the panel data analysis arises from restricting the potential outcomes. Assumptions \ref{ass:twfe} and \ref{ass:constant} do exactly that by imposing a very special structure on the underlying potential outcomes. Both of these assumptions are restrictive, and the new causal panel data literature discussed in this section focused on fully relaxing Assumption \ref{ass:constant} while maintaining the two-way model for a particular set of potential outcomes that are effectively used as a control group.

In practice, the choice of the control group depends on application details and, to a certain degree, is arbitrary. From this perspective, the analysis that fully relaxes Assumption \ref{ass:constant} while keeping Assumption \ref{ass:twfe} is somewhat internally inconsistent. If we are willing to assume that the baseline potential outcomes follow a simple model, then it is not clear why we would not be willing to make a similar assumption for the treatment effects. After all, the differences in the baseline potential outcomes partly arise due to other unobserved treatments, and if their effect is fully heterogeneous, then the two-way model is unlikely to hold. 

One interpretation of the results in this section is that we do not need to take a stand on the degree of treatment effect heterogeneity because there exist methods that are fully robust to it as long as the two-way model holds, allowing us to relax one assumption at a time. This conclusion, however, is unlikely to hold in more complicated settings, \textit{e.g.}, see \cite{arellano2001panel} for a related impossibility result in sequentially exogenous models. One can still attempt to relax Assumption \ref{ass:constant} but simultaneously relax at least part of  Assumption \ref{ass:twfe}. Which one of those approaches deserves more attention is fundamentally an empirical question.  More broadly, we recommend letting the data determine 
 the degree of underlying heterogeneity in potential outcomes. In the next section, we discuss a class of methods that does exactly that.  

\section{Moving Away from the Two-Way Fixed Effect Structure}\label{section:relax}
A key strand of the recent causal panel literature starts with the introduction of the Synthetic Control (SC) method by Alberto Abadie and coauthors, initially in    \citep{abadie2003}, with more detailed methodological discussions in  \citep*{abadie2010synthetic} and \citep{abadie2014}. This brought a substantially different perspective to the questions studied in the TWFE literature. Initially, the SC literature remained very separate from the  TWFE discussions. The SC literature focused on imputing missing potential outcomes by creating synthetic versions of the treated units constructed as convex combinations of control units. This more algorithmic, as opposed to model-based, approach has inspired much new research, ranging from factor-model approaches that motivate synthetic-control type algorithms to hybrid approaches that link synthetic control methods to the earlier TWFE methods and highlight their connections.

In this section we first discuss the basic synthetic control method in Section \ref{section:synth_basics}. Next, in Section \ref{section:matrix}, we discuss the direct estimation of factor models. In Section \ref{section:hybrid} we discuss some hybrid methods that combine synthetic control and TWFE components.

\subsection{Synthetic Control Methods}\label{section:synth_basics}

In the original paper, \citep{abadie2003} were interested in estimating the causal effect on terrorism on the Basque region economy. They constructed a comparison for the Basque region based on a convex combination of other regions in Spain. The weights were chosen to ensure that this synthetic Basque region matched the actual Basque region closely in the years pre-treatment (prior to the terrorism) years.

In a short period of time, this synthetic control method has become a very widely used approach, popular in empirical work in social sciences, as well as in the popular press (including The Economist and The Guardian), with many theoretical advances in econometrics, statistics, and computer science. The key papers by Abadie, Diamond, and Hainmueller that discuss the details of the original synthetic control proposals are  \citep*{abadie2010synthetic, abadie2014}. For  recent reviews see \citep{abadie2019using} and \citep{samartsidis2019assessing}.

\subsubsection{Estimation}

Here we use a  characterization of the SC method as a least squares estimator, as discussed in
 \citep{doudchenko2016balancing}, that is slightly different from that in  \citep*{abadie2010synthetic}. 
We   focus on the case without covariates. 
 Suppose unit $N$ is the sole treated unit, and is treated in period $T$ only. Define the weights $\hat\omega$ as the regression estimates subject to restrictions:
 \begin{equation}\label{eq:sc}\hat\omega\equiv \arg\min_{\omega|\omega\geq 0,\sum_j\omega_j=1}\sum_{t=1}^{T-1} \left(Y_{Nt}-\sum_{j=1}^{N-1} \omega_j Y_{jt}\right)^2,\end{equation}
and then impute the missing potential outcome as
\[\hat Y_{NT}(0)=\sum_{j=1}^{N-1} \hat\omega_j Y_{jT}.\]
The nonnegative weights $\hat\omega_j$ define the ``synthetic'' control that gave the methods its name.
One remarkable finding in the initial papers by Abadie and coauthors is that this solution is typically sparse, with positive weights $\hat\omega_j>0$ typically only for a small subset of the control units. Although this is not always important substantively, it greatly facilitates the interpretation of the results. For example, in the German reunification application in \citep{abadie2014} where the full set of potential controls consists of sixteen OECD countries, only five countries, Austria, Japan, The Netherlands, Switzerland, and the US, have positive weights.

The characterization of the SC estimator in (\ref{eq:sc}) allows for an interesting comparison with methods based on the unconfoundedness assumption discussed in Section \ref{section_unconf}. With a linear model specification, unconfoundedness would suggest an estimator
\begin{equation}\label{eq:unc}\hat\beta=\arg\min_{\beta}\sum_{i=1}^{N-1} \left(Y_{iT}-\beta_0-\sum_{s=1}^{T-1} \beta_s Y_{is}\right)^2,\end{equation}
followed by the imputation of the missing potential outcome as
\[\hat Y_{NT}(0)=\hat\beta_0+\sum_{s=1}^{T-1} \hat\beta_s Y_{Ns}.\]
The difference is that in, using the terminology of \citep*{athey2021matrix}, SC in the regression in (\ref{eq:sc}) relies on \emph{vertical regression} with $T-1$ observations and $N-1$ predictors, with some restrictions on the parameters  (with the units of observations corresponding to the columns of $\by$), and (\ref{eq:unc}) relies on \emph{horizontal regression} with $N-1$ observations and $T$ regressors after including an intercept (with the units of observations corresponding to the columns of $\by$).
If the minimizers in these least squares regressions are not unique, we take the solution to be the one that minimizes the $L_2$ norm (\citealp{spiess2023double}). See \citep*{shen2022tale} for more insights into the comparison between the horizontal and vertical regressions in this setting.  In particular, they demonstrate the interesting insight that point estimates for the counterfactuals are identical for the vertical and horizontal regressions in the absence of the nonnegativity and adding up restrictions.

One interesting aspect of the synthetic control approach is that it is more algorithmic than many other methods used in these settings. Consider the estimator based on unconfoundedness in (\ref{eq:unc}). Such an approach is typically motivated by a linear model
\[ Y_{iT}=\gamma_0+\sum_{s=1}^{T-1} \gamma_s Y_{is}+\varepsilon_{i},\]
with assumptions on the $\varepsilon_i$ given the lagged outcomes. The corresponding model for the SC estimator would be
\[ Y_{Nt}=\sum_{j=1}^{N-1} \omega_j Y_{jt}+\eta_{t},\]
with assumptions on $\eta_t$ given the contemporaneous outcomes for other units. However, such assumptions are rarely postulated, and for good reason. It would postulate a relationship between the cross-section units, {\it e.g.,} states, that is oddly asymmetric. If, as in the application in 
\citep*{abadie2010synthetic}, California is the treated state, this model would postulate a relationship between California and the other states of a form that cannot also hold for all other states. Attempts to specify models that justify the synthetic control estimator had limited success. 
\citep*{abadie2010synthetic} discusses factor models as a data generating process, but that begs the question of why one would not directly estimate the factor model.
Researchers have done so, as discussed in Section \ref{section:matrix} below, but interestingly, such attempts have not always outperformed the Abadie-Diamond-Haimueller synthetic control methods, suggesting the latter have attractive properties that are not fully understood yet. See the review in \citep{abadie2019using} for more discussion on conditions under which synthetic control methods are appropriate.

In \cite{arkhangelsky2024sequential}, the authors show that in environments where the contribution of idiosyncratic errors to the overall variation in the outcomes is small, {\it i.e.}, most of the variation is explained by the two-way fixed effects and the factor model, a version of the SDID estimator discussed in Section \ref{section:hybrid} is asymptotically equivalent to a particular methods-of-moments estimator for the underlying factor model.  These results are relevant for applications where researchers rely on aggregated data, such as the GRCS data discussed in Section \ref{grcs}, and bridge the gap between methods that directly estimate factors models and those based on the SC ideas. Further theoretical research is needed to better understand this connection in other settings.

\subsubsection{Modifications}

A number of modifications have been suggested to the basic version of the SC estimator.
\citep{hsiao2012panel, doudchenko2016balancing} and \citep{ferman2021synthetic} suggest making the estimator more flexible by allowing for an intercept in the regression (or, equivalently, applying the method to outcomes in deviations from time-averages). 
 \citep{hsiao2012panel, doudchenko2016balancing, gardeazabal2017empirical} also discuss allowing the weights to be outside the unit interval. This improves the in-sample fit but has the potential of making the out-of-sample predictions less accurate. 

\citep{li2023frontiers}
proposes an alternative to TWFE estimation that relies on selecting a set of controls. One can think of this as a special case of SC where the weights for the control units are either zero, or $1/N_C$, where $N_C$ is the number of control units selected. The proposal includes a greedy algorithm for selecting the set of controls with an objective function that closely mimics the SC criterion for the case with an intercept.

Typically, in the synthetic control method, only the control units are weighted. In principle, however, one could also weight the treated units to make it easier to find a set of (weighted)  control units that are similar to these weighted treated units during the pre-treatment period, as suggested in 
\citep{kuosmanen2021design}.

\citep{kellogg2021combining}
suggest combining matching and synthetic control methods. Whereas synthetic control methods avoid extrapolation at any cost, combining it with matching allows researchers to lower the bias from either method. 

\subsubsection{Regularization}

In settings where the number of control units is large relative to the number of pre-treatment periods, this requires some form of regularization. \citep{hsiao2012panel} use statistical information criteria.
\citep{doudchenko2016balancing} suggest regularizing the weights by imposing  an elastic net penalty on the weights $\omega_i$, with the penalty chosen by cross-validation.
\citep{spiess2023double} avoid the choice of a penalty term by  choosing the minimum $L_2$ norm value for the weights within the set of weight combinations that lead to the optimal in-sample fit, in the spirit of the recent double descent literature (\citealp{belkin2019reconciling}).
\citep{abadie2021penalized} recognizes that weights that a convex combination of control units that are all far away from the treated unit are not as attractive as a convex combination of control units that are all close to the target treated unit. They suggest choosing the weights by minimizing the sum of the original synthetic control criterium and a term that penalizes the distance between any of the control units and the target unit
\[\hat\omega=\arg\min_{\omega|\omega\geq 0,\sum_j\omega_j=1}\sum_{t=1}^{T-1} \left(Y_{Nt}-\sum_{j=1}^{N-1} \omega_j Y_{jt}\right)^2+\lambda\sum_{j=1}^{N-1} \omega_j \sum_{t=1}^{T-1} (Y_{Nt}-Y_{jt})^2,
\]
with the tuning parameter $\lambda$ chosen through cross-validation, for example, on the control units.

\subsubsection{Inference}

Inference has been a major challenge in synthetic control settings, and there is, as of yet, no consensus regarding the best way to estimate variances or construct confidence intervals. One particular challenge is that the methods are often used in settings with just a single treated unit/period, or relatively few treated unit/period pairs, making it difficult to rely on central limit theorems for the distribution of estimators. In applications where the number of units and periods is large, the situation is different; see the results in \cite{arkhangelsky2021synthetic, ferman2021properties}.

One approach has been to use placebo methods to test sharp null hypotheses, typically for the null hypothesis of no effect of the intervention.
\citep*{abadie2010synthetic} proposes such a method.
Suppose there is a single treated unit, say unit $N$. 
\citep*{abadie2010synthetic} construct a distribution of estimates based on each control unit being analyzed as the treated unit and then calculate the p-value for unit $N$ as the quantile in that distribution of placebo estimates. See also \cite{firpo2018synthetic} for an extension and additional analysis of this method.

\citep*{doudchenko2016balancing} suggests that the same placebo approach can be based on changing the time period that was treated. Essentially here the idea is to think of the time of the treatment as random, generating a randomization distribution of estimates.
In a related approach \citep*{chernozhukov2021exact, viviano2023synthetic, lei2021conformal} 
develop conformal inference procedures that rely on the exchangeability of the residuals from some model over time.
 \citep*{cattaneo2021prediction}  propose the construction of prediction intervals for the counterfactual outcome.

\subsection{Matrix Completion  Methods and Factor Models}\label{section:matrix}

A second set of methods that relaxes the TWFE assumptions focuses directly on factor models, where the outcome is assumed to have the form
\begin{equation}\label{factor} Y_{it}(0)=\sum_{r=1}^R \alpha_{ir}\beta_{tr}+\varepsilon_{it}.\end{equation}
First, note that this generalizes the TWFE specification: if we fix the rank at $R=2$, and set $\alpha_{i2}=1$ for all $i$ and $\beta_{t1}=1$ for all $t$, this is identical to the TWFE specification, but the factor model obviously allows for more general dependence structures in the data.
Although such factor models have a long tradition in panel data, {\it e.g.,} \citep{anderson1984estimating,chamberlain1983arbitrage,stock1998diffusion, bai2002determining,bai2009panel}, the recent causal literature has used them in different ways.

\subsubsection{Matrix Completion with Nuclear Norm Regularization}

\citep*{athey2021matrix} take an approach that models the entire matrix of potential control outcomes as
\[ Y_{it}(0)=L_{it}+\alpha_i+\beta_t+\varepsilon_{it},\]
where the $\varepsilon_{it}$ is random noise, uncorrelated with the other components. The matrix $\mathbf{L}$ with typical element $L_{it}$ is a low-rank matrix. As mentioned above the unit and time components $\alpha_i$ and $\beta_t$ could be subsumed in the low-rank component as they on their own form a rank-two matrix, but in practice it improves the performance of the estimator substantially to keep these fixed effect components in the specification separately from the low-rank component $\mathbf{L}$. The reason is that we regularize the low rank component $\mathbf{L}$, but not the individual and time components.
Building on the matrix completion literature (\citep{candes2009exact, candes2010matrix}), \citep*{athey2021matrix} propose the Nuclear-Norma-Matrix-Completion (NNMC) estimator based on minimizing
\[
\sum_{i=1}^N\sum_{t=1}^T (1-W_{it})
\left(Y_{it}-L_{it}-\alpha_i-\beta_t
\right)^2+\lambda\|\mathbf{L}\|_*,
\]
over $\mathbf{L}$, $\alpha$, and $\beta$. The missing $Y_{it}(0$ values are then imputed using the estimated parameters.
Here the nuclear norm $\|\mathbf{L}\|_*$ is the sum of the singular values $\sigma_l(\mathbf{L})$ of the matrix $\mathbf{L}$, based on the singular value decomposition $\mathbf{L}=\mathbf{S}\Sigma\mathbf{R}$, 
where $\mathbf{S}$ is $N\times N$, $\Sigma$ is the $N\times T$ diagonal matrix with the singular values and 
$\mathbf{R}$ is $T\times T$. The penalty parameter
$\lambda$ is chosen through out-of-sample crossvalidation.  
The nuclear norm regularization shrinks towards a low rank estimator for $\mathbf{L}$, similar to the way LASSO shrinks towards a sparse solution in linear regression.

\subsubsection{Robust Synthetic Control}

\citep*{amjad2018robust}  focus on the case with a single treated unit. 
They start with a factor model $\mathbf{Y}=\mathbf{L}+\varepsilon$. They would like to use a synthetic control estimator with denoised matrix $\mathbf{L}$ as the control outcomes, rather than the actual outcomes $\mathbf{Y}$. They implement this through a two step procedure. In the first step the matrix  $\mathbf{L}$ is estimated by taking the singular value decomposition, and setting all singular values below a threshold $\mu$ equal to zero. This leads to a low-rank estimate $\hat{\mathbf{L}}$, which is then scaled by one over $p$, where $p$ is the maximum of the fraction of observed outcomes and $1/((N-1)T)$. 

In the second step \citep*{amjad2018robust} use the part of this rescaled matrix corresponding to the control units, in combination with the pre-treatment-period values of for treated unit, in a standard synthetic control approach. The idea is that using de-noised outcomes $\hat{\mathbf{L}}$ instead of the actual  outcomes $\mathbf{Y}$ leads to better predictors by removing an estimate of  the noise component $\varepsilon$. In this second synthetic control step \citep*{amjad2018robust} do not impose the convexity restrictions on the weights, but do add a regularization penalty.

\subsubsection{Interactive Fixed Effect or Factor  Models}

Building on the factor model literature in econometrics (\citealp*{chamberlain1983arbitrage,holtz1988estimating,chamberlain1992efficiency, pesaran2006estimation, bai2009panel, moon2015linear, moon2018nuclear,freyberger2018non}), 
\citep{xu2017generalized}
studied direct estimation of factor models as an alternative to synthetic control methods. 
The basic setup models the control potential outcome as in (\ref{factor}).
The number of factor is then estimated or pre-specified and the model is directly estimated after some normalization. Based on this model on can impute the missing potential outcomes for the treated unit/time-period pairs and use that to estimate the average effect for the treated.
See \citep*{gobillon2016regional} for an application.

\subsubsection{Grouped Panel Data}

\citep{bonhomme2015grouped, bonhomme2022discretizing}
consider a factor model but impose a group structure. In our causal setting, their setup would correspond to
\[ Y_{it}(0)=\theta_{G_i,t}+\varepsilon_{it},\]
with the group membership unknown. They focus on the case with the number of groups $G$ known.
In that case one can write the model as a factor model with $G$ factors $\lambda_{rt}$ and the loadings equal to indicators, $\alpha_{ir}=\mathbf{1}_{G_i=r}$, so that
\[ Y_{it}(0)=\theta_{G_i,t}+\varepsilon_{it}
=\sum_{r=1}^G \alpha_{ir}\lambda_{rt}+\varepsilon_{it}.\]
Computationally this grouped structure creates substantial challenges.  \citep{mugnier2022make, chetverikov2022spectral} suggest alternative estimation methods that are computationally more attractive. 

\subsubsection{Tuning}

One disadvantage the methods discussed in this section share is the need to specify the tuning parameters. This sets them apart from the conventional TWFE methods we discussed before and makes them harder to adopt in practice. In the case of the matrix completion estimator proposed by \citep*{athey2021matrix}, this tuning parameter is the regularization parameter $\lambda$ that quantifies the importance of the nuclear norm penalty. In the context of the standard interactive fixed effects estimators, one needs to specify the rank of the underlying factor model. The same applies to the estimator based on finitely many groups. In principle, one can use traditional techniques from the machine learning literature, such as cross-validation, to find appropriate values of these parameters. The panel dimension, however, creates an additional challenge on how exactly to implement the cross-validation.  It is thus attractive to have methods that generalize the two-way methodology and do not require explicit tuning. One such proposal is \citep{moon2018nuclear}, where the authors analyze the limiting version of the estimator from \citep*{athey2021matrix} with $\lambda$ approaching zero. They show that the resulting estimator is consistent under relatively weak assumptions, albeit it can converge at a slower rate.

\subsection{Hybrid Methods}\label{section:hybrid}

Two recent methods combine some of the benefits from the synthetic control approach with either TWFE ideas or with unconfoundedness methods. These methods are particularly attractive because the nest TWFE, while being able to accomodate more flexible outcome models. There are in essence two approaches. One can directly generalize the outcome model, or one can use a local version of the TWFE model. This is somewhat  similar to the way one can generalize a linear regression model by making the regression function more flexible through the inclusion of additional function of the regressors, or by estimating it locally through kernel methods.

\subsubsection{Synthetic Difference In Differences}

For expositional reasons let us consider the case with a single treated unit and time period, say unit $N$ in period $T$, although the insights readily extend to the block assignment case.
Once the researcher has calculated the SC weights, the SC estimator for the treatment effect can be characterized as a weighted least squares regression,
\begin{equation}\label{eq:screg} \min_{\beta,\tau}\sum_{i=1}^N\sum_{t=1}^T \hat\omega_i\left(Y_{it}-\beta_t-\tau W_{it}\right)^2.\end{equation}
It is useful to contrast this with the TWFE estimator, which is based on a slightly different least squares regression:
\begin{equation}\label{eq:didreg}  \min_{\beta,\alpha,\tau}\sum_{i=1}^N\sum_{t=1}^T \left(Y_{it}-\alpha_i-\beta_t-\tau W_{it}\right)^2.\end{equation}
The two differences are that  \begin{inparadesc}\item[$(i)$], the SC regression in (\ref{eq:screg}) uses weights $\hat\omega_i$, and \item[$(ii)$] the TWFE regression in (\ref{eq:didreg}) has unit-specific fixed effects $\alpha_i$.
\end{inparadesc}

In the light of
of this comparison, and more generally in the context of the larger panel data literature, the omission of the unit fixed effects from the synthetic control regression may seem surprising. 
\citep*{arkhangelsky2021synthetic} exploit this  by proposing what they call the Synthetic Difference In Difference (SDID) estimator that includes both the unit fixed effects $\alpha_i$ and the SC weights $\hat\omega_i$, as well as analogous time weights $\hat\lambda_t$, leading to 
\[ \min_{\beta,\alpha,\tau}
\sum_{i=1}^N\sum_{t=1}^T \hat\omega_i\hat\lambda_t\left(Y_{it}-\alpha_i-\beta_t-\tau W_{it}\right)^2.\]
 The time weights $\hat\lambda_t$ are calculated in a way similar to the unit weights,
\[ \min_\lambda \sum_{i=1}^{N-1}
\left( Y_{iT}-\sum_{s=1}^{T-1}\lambda_s Y_{is}\right)^2,\]
subjet to the restriction that $\lambda_s\geq 0$, and $\sum_{s=1}^{T-1} \lambda_s=1$. The weights for treated units and periods are equal to $1$.

\subsubsection{Augmented Synthetic Control}

\citep{ben2021augmented} augment the SC estimator by regressing the outcomes in the treatment period on the lagged outcomes using data for the control units. Suppose that, following \citep{ben2021augmented} one uses ridge regression for this first step, again in the setting with unit $N$ and period $T$ the only treated unit/time-period pair:
\[ \hat\eta=\arg\min_\eta \sum_{i=1}^{N-1} \left( Y_{iT}-\eta_0-\sum_{s=1}^{T-1} \eta_s Y_{is}\right)^2+\lambda \sum_{s=1}^{T-1} \eta_s^2,\]
with ridge parameter $\lambda$ chosen through cross-validation.
A standard unconfoundedness approach would predict the potential control outcome for the treated unit/time period pair as
\[  \hat Y_{NT}=\hat\eta_0+\sum_{s=1}^{T-1}
\hat\eta_s Y_{Ns}
.\]
The augmented SC estimator modifies this by combining it with SC weights in a way that can be seen either as a bias-adjustment to the unconfoundedness estimator, or a bias-adjustment to the SC estimator:
\[ \hat Y_{NT}=
\hat\eta_0+\sum_{s=1}^{T-1}
\hat\eta_s Y_{Ns}+
\sum_{i=1}^{N-1} \omega_i\left( Y_{iT}-\hat\eta_0-\sum_{s=1}^{T-1}
\hat\eta_s  Y_{is}\right)
\]
\[ =\sum_{i=1}^{N-1} \omega_i Y_{iT}+\sum_{s=1}^{T-1}
\hat\eta_s \left(Y_{Ns}-\sum_{j=1}^{N-1} \omega_j Y_{js}
\right).\]
\citep*{ben2022synthetic} extend this approach to the case with staggered adoption.

\subsubsection{The Connection Between Unconfoundedness, Difference-In-Differences, Synthetic Control and Matrix Completion}

Although methods based on unconfoundedness, and synthetic control estimators, difference-in-differences, and matrix completion estimators appear to be quite different, they are in fact closely related.  We want to highlight two insights regarding these connections.

We focus on the case with a single treated unit/period pair, say unit $N$ in period $T$,	The observed control outcomes are $\by$, an $N\times T$ matrix with the $(N,T)$ entry missing. We  partition this matrix as
\[ \by=\left(
\begin{array}{cc}
	\tty & \bye\\
	\bytt & ?
	\end{array}\right),\]
	where $\tty$ is a $(N-1)\times (T-1)$ matrix, and
	$\bye$ and $\byt$ are  $(N-1)$
and $(T-1)$ component vectors, respectively.

First, \citep*{shen2022tale} discuss an interesting connection between SC estimators and estimators based on unconfoundedness in combination with linearity. In that case we first estimate a linear regression
\[ Y_{iT}=\gamma_0+\sum_{s=1}^{T-1} \gamma_s Y_{is}+\varepsilon_{i},\]
and then impute the missing outcome as $\hat{Y}_{NT}=\hat\gamma_0+
\sum_{s=1}^{T-1} \hat\gamma_s Y_{Ns}$. \citep*{shen2022tale} show that if we drop the intercept from this regression, $\gamma_0=0$,  then the unconfoundedness imputation is identical to the SC imputation (where we also restrict the weights to be nonnegative and sum to one).

In an alternative connection,
\citep*{athey2021matrix} show that in some cases linear versions of all four estimators can all be characterized as solutions to the same optimization problem, subject to different restrictions on parameters of that optimization problem.
To see this, 
define for a given positive integer $R$, an $N\times R$ matrix $\bu$, an $T\times R$ matrix $\bv$, an $N$-vector $\alpha$ and a $T$-vector $\beta$, and a scalar $\lambda$ the objective  function
\begin{equation}\label{obj}
Q(R,\bu,\bv,\alpha,\beta,\lambda)\equiv
\sum_{i=1}^N\sum_{t=1}^T
(1-W_{it})\left( Y_{it}-
\sum_{r=1}^R U_{ir}V_{tr}-\alpha_i-\beta_t
\right)^2
		\end{equation}
  \[+\lambda\left(\sum_{i=1}^N\sum_{r=1}^R U_{ir}^2+\sum_{t=1}^T\sum_{r=1}^R V_{tr}^2\right).\]
When $R=0$, we take the product $\bu\bv^\top$ to be the $N\times T$ matrix with all elements equal to zero.
Given $\bu$, $\bv$, $\alpha$ and $\beta$ the imputed value for $Y_{NT}$ is
$\hat{Y}_{NT}=\sum_{r=1}^R U_{Nr}V_{Tr}-\alpha_i-\beta_t$.

First note that  minimizing  the objective function (\ref{obj}) over  the rank $R$, the matrices $\bu$, $\bv$ and the  vectors $\alpha$ and $\beta$ given $\lambda=0$, does not lead to a unique solution. By choosing the rank $R$ to the minimum of $N$ and $T$, we can find for any pair $\alpha$ and $\beta$ a solution for $\bu$ and $\bv$ such that
$(1-W_{it})( Y_{it}-
\sum_{r=1}^R U_{ir}V_{tr}-\gamma_i-\delta_t
)=0$ for all $(i,t)$, with different imputed values for $Y_{NT}$.
The implication is that we need to add some structure to the optimization problem. The next result shows that unconfoundedness regression,  the  SC estimator, the DID estimator, and the MC estimator can all be expressed as minimizing  the objective function under different restrictions on,  or with different approaches to regularization of, $(R,\bu,\bv,\alpha,\beta)$.

\noindent{\sc Nuclear Norm Matrix Completion}
The nuclear norm matrix completion estimator chooses $\lambda$ through cross-validation.

\noindent{\sc Unconfoundedness}
The unconfoundedness regression is based on regressing $\by_1$ on $\bytt$ and an intercept. It can also be characterized as the solution to minimizing (\ref{obj})
with
the restrictions
\[
		R=T-1,\hskip0.5cm
		\bu=
		\left(
		\begin{array}{c}
		\tty \\
		\bytt
		\end{array}\right),\quad \alpha=0, \quad \beta_1=\beta_2=\ldots=\beta_{T-1}=0,\qquad \lambda=0.
		\]

\noindent{\sc Synthetic Control}
  The SC estimator imposes the restrictions
subject  to
\[R=N-1, \quad \bv=
		\left(
		\begin{array}{cc}
		\mathbf{Y}_0^\top \\
		\bye^\top
		\end{array}\right),\quad
		\alpha=0, \hskip0.5cm \beta=0, \quad
		\forall\ i, U_{iT}\geq 0,\ \sum_{i=1}^{N-1} U_{iT}=1,\quad \lambda=0.
		\]

\noindent{\sc Difference In Differences}
  The DID estimator fixes $R=0$.

\subsubsection{The Selection Mechanism}\label{subsec:selection}

Another set of insights concerning the differences between the various estimators emerges from a focus on the selection mechanism. First, \cite*{ghanem2022selection} show that outside of a few special cases, to justify the conventional parallel trends assumption, one needs to assume that the treatment is unrelated to time-varying components of the outcomes. The restrictiveness of this assumption presents a challenge for the DID estimator, which relies on this  parallel trends assumption. These concerns are less relevant for other estimators; see \citep{arkhangelsky2023synth, viviano} for two recent discussions.

We can see this in the setting with the block design, where there are $T_0+1$ periods, and some units are treated in the last period, with $D_i$ being the treatment indicator, $D_i=\mathbf{1}_{W_{iT}=1}$. Suppose the underlying potential outcomes follow a static two-way model of Section \ref{section:twe}:
\begin{equation*}
    Y_{it}(0) = \alpha_i + \beta_t + \varepsilon_{it},\quad  \varepsilon_{it} \indep \alpha_i, \quad \tau = Y_{it}(1) - Y_{it}(0).
\end{equation*}
The key feature that determines the performance of different algorithms in this environment is the relationship between $D_i$ and the vector of errors $(\varepsilon_{i1},\dots, \varepsilon_{iT_0+1})$. As long as $D_i$ is mean-independent from $(\varepsilon_{i1},\dots, \varepsilon_{iT_0+1})$, then the discussed estimators will have good statistical properties. This should not be surprising for the DID (which does not rely on large $T_0$) or matrix completion estimator because their statistical properties are established under this assumption. The fact that the SC estimator would work well in this situation follows from the results in \citep*{arkhangelsky2021synthetic} and \citep*{arkhangelsky2023synth}. 

This conclusion changes dramatically if we allow $(\varepsilon_{i1},\dots, \varepsilon_{iT_0+1})$ to be correlated with $D_i$. If this correlation is completely unrestricted, then any observed differences in outcomes in the two groups can be attributed to differences in errors, and it is impossible to identify the effect using any method. Suppose, however, that we make a natural selection assumption
\begin{equation*}
    D_i \indep Y_{iT_0+1}(0) | \alpha_i, Y_{i1}, \dots, Y_{iT_0}(0), 
\end{equation*}
which restricts the correlation of $D_i$ with $\varepsilon_{i,T_0+1}$. Note that this restriction combines both the selection on fixed effect assumption discussed in Section \ref{sec:par_trends} and the unconfoundedness assumption discussed in Section \ref{section_unconf}. 

As long as $\varepsilon_{it}$ are autocorrelated, the DID estimator is inconsistent, even when $T_0$ goes to infinity. The reason for this failure is that $\varepsilon_{iT_0+1} - \sum_{t\le T_0} \varepsilon_{it}/T_0$ remains correlated with $D_i$ which introduces bias. The performance of the SC estimator is different, and the results in \citep*{arkhangelsky2023synth} show that the SC estimator is consistent and asymptotically unbiased as long as $T_0$ goes to infinity. 
The consistency properties of the matrix completion estimator and the unconfoundedness regression are not established for this setting.

\citep{viviano} focuses on a factor model with block assignment where $D_i$ can be correlated with the factor loadings and the time of initial exposure can be correlated with the factors. They present conditions under which the SC estimator is consistent.

This discussion illustrates that to analyze the behavior of algorithmically related estimators one needs to take a stand on the underlying selection mechanism. Most of the recent results in the causal panel data literature are established under strict exogeneity, which does not allow $D_i$ to be correlated with $\varepsilon_{it}$. Understanding the performance of different estimators in environments where such correlation is present is an attractive area of future research that can benefit from the econometric panel data literature.

\subsubsection{Simulation Comparisons}

There have been a number of studies comparing various DID and SC  estimators in simulations.
These have not always been in realistic settings, limiting their usefulness for practitioners.
In fact, it is not as easy in longitudinal setting to come up with realistic simulation settings that capture both the degree of cross-section and time-series dependence that is present in a given data set. In pure cross-section settings, \cite{athey2021using} suggests using generative adversarial networks to generate realistic data-generating processes, but that does not immediately extend to longitudinal settings.
\cite{arkhangelsky2021synthetic}
compare DID, SC, SDID, and NNMC estimators in settings motivated by state and country panel data sets. They first estimate factor models with four factors and use that to construct a data-generating process with additive fixed effects, four additional factors, and an autoregressive error process. They find that there is substantial variation in the performance of the methods, with SDID typically outperforming the other methods.

\section{Nonlinear Models}\label{section:nonlinear}

In this section, we discuss some nonlinear panel data models. By nonlinear models, we mean here models where the conditional mean function is not linear in parameters. Part of this literature is motivated by the concern that the standard fixed effect models maintain additivity and linearity in a way that does not do justice to the type of data that are often analyzed. With binary outcomes, it is particularly difficult to justify the standard TWFE model. At the same time, estimating the unit and time fixed effects inside a logistic or probit model does not lead to consistent estimators for the effects of interest in typical settings.

\subsection{Changes-In-Changes}\label{cic}

\citep*{athey2006identification} focus on the repeated cross-section case with two periods and two groups, one treated in the second period and one never treated. They are concerned with the functional-form dependence of the standard TWFE specification in levels.
If the model \[ Y_{i}(0)=\mu+\alpha \mathbf{1}_{C_i=1}+\beta\mathbf{1}_{T_i=1}+\varepsilon_i,\]
holds in levels, then obviously it cannot hold in general in logarithms. In fact, in some cases, one can test that the model cannot hold in levels. Suppose the outcome is binary, and suppose that the potential control outcome averages by group and time period are $\overline{Y}_{11}(0)=0.2$ (for the first-period control group), $\overline{Y}_{12}=0.8$ (for the second-period control group), and $\overline{Y}_{21}(0)=0.7$ (for the first-period treatment group). Then the additive TWFE model implies that the second-period treatment group, in the absence of the treatment, would have had average outcome 0.7+(0.8-0.2)=1.3, which of course, is not feasible with binary outcomes.

To address this concern
\citep*{athey2006identification} propose a scale-free changes-in-changes (CIC) model for the potential control outcomes,
\[ Y_i=g(U_i,T_i),\]
where the $U_i$ is an unobserved component that has a different distribution in the treatment group and the control group but a distribution that does not change over time. The standard TWFE model can be viewed as the special case where $g(u,t)$ is additively separable  in $u$ and $t$:
\[ g(u,t)=\beta_0+u+\beta_1 t,\]
implying that the expected control outcomes can be written  in the TWFE form  as
\[ \mme[Y_i(0)|T_i=t,G_i=g)=\beta_0+\beta_1 t+\alpha \mme[U_i|G_i=g]-\mme[U_i|G_i=1].\]
\citep*{athey2006identification} show that if $U_i$ is a scalar, and $g(u,t)$ is strictly monotone in $u$, one can infer the second-period distribution of the control potential outcome in the treatment group as
\[ F_{Y_i(0)|T_i=2,G_i=2}(y)=F_{Y_i(0)|T_i=1,G_i=2}\left(F^{-1}_{Y_i(0)|T_i=1,G_i=1}\left(F_{Y_i(0)|T_i=2,G_i=1}(y)\right)\right).\]
This, in turn, can be used to estimate the average effect of the intervention on the second-period outcomes for the treatment group.

The expression for the counterfactual distribution of the control outcome for the second-period treatment group has an analog in the literature on wage decompositions, see \citep*{altonji1999race}. \citep*{arkhangelsky2019dealing} discusses a similar approach to the CIC estimator in \citep*{athey2006identification}, where the role of the groups and time periods are reversed and also considers an extension for multiple outcomes. \citep{wooldridge2022simple} also studies nonlinear versions of DID/TWFE approaches. In the two-period-two-group setting, his starting point assumes there is a known function $g:\mathbb{R}\mapsto\mathbb{R}$ such 
that $\mme[Y_{it}(0)|D_{i}]=g(\mu+\alpha D+\gamma_t)$, so that there is a parallel trend inside the known transformation $g(\cdot).$  The transformation $g(\cdot)$ could be the exponential function, $g(a)=\exp(a)$ in case of non-negative outcomes, or the logistic function $g(a)=\exp(a)/(1+\exp(a))$ in case of binary outcomes.

\subsection{Distributional Synthetic Controls}

\citep{gunsilius2023distributional} develops a model that has similarities to both the CIC and SC control approaches. He focuses on a setting with repeated cross-sections, where we have a relatively large number of units observed in a modest number of groups, with a modest number of time periods.
As in the canonical synthetic control case there is a single treated group. 
Whereas the synthetic control method chooses weights on the control units so that the weighted controls match the treated outcomes in the pre-treatment periods, 
the \citep{gunsilius2023distributional} approach chooses weights on the control groups so that the marginal distribution for the weighted controls matches that for the treated group. The metric is based on the quantile function $F_{Y_{gt}}^{-1}(v)$,  for group $g$ and period $t$.
First, weights $\hat\omega_{tg}$ are calculated separately for each pre-treatment period $t$ 
based on the
following objective:
\[ \hat\omega_{tg}=\arg\min_{\omega:\omega\geq 0, \sum_{g=1}^{G-1}\omega_{gt}=1}
\frac{1}{M}\sum_{m=1}^M
\left(
\sum_{g=1}^{G-1} \omega_{gt}\hat F_{Y_{gt}}^{-1} (V_m)-
\hat F_{Y_{Gt}}^{-1} (V_m)
\right)^2,
\]
where the quantile functions are evaluated at $M$ randomly choosing values $v_1,\ldots,v_M$.

In the next step the weights are averaged over time,
\[ \hat\omega_g=\frac{1}{T-1}\sum_{t=1}^{T-1} \hat\omega_{gt}.\]
Finally the quantile function for the treated group in the absence of the treatment is estimated as the synthetic control average of the control quantile functions:
\[ \hat F^{-1}_{GT}(v)=\sum_{g=1}^{G-1} \hat\omega_g
\hat F^{-1}_{gT}(v).\]
Note that in the case with $G=2$, so there is just a single control group, the quantile function for the treated group in the last period in the absence of the treatment is identical to the quantile function for the control group in the last period, and the pre-treatment distributions are immaterial.

\subsection{Balancing Statistics to Control for Unit Differences}

\citep*{arkhangelsky2022doubly} focus on settings where the treatment can switch on and off, as in the assignment matrix in Equation (\ref{eq:general}),  unlike the staggered adoption case where the treatment can only switch on. They also assume there are no dynamic effects. Their focus is on flexibly adjusting for differences beyond additive effects. Allowing for completely unrestricted differences between units would require relying solely on within-unit comparisons. Often the number of time periods is not sufficient to rely on such comparisons and still obtain precise estimates. \citep*{arkhangelsky2022doubly} balance these two concerns, the restrictiveness of the TWFE model and the lack of precision when focusing purely on within-unit comparisons, by making assumptions that allow the between-unit differences to be captured by a low-dimensional vector, which then can be adjusted for in a flexible, nonlinear way using some of the insights from the cross-section causal inference literature.

To see the insights most clearly it is useful to start with a simpler setting. Specifically, let us first consider a clustered sampling setting with cross-section data studied in \citep*{arkhangelsky2018role}. In that case a common approach is based on a fixed effect specification
\[ Y_i=\alpha_{C_i}+\tau W_i+\beta X_i+\varepsilon_i,\]
where $C_i$ is the cluster indicator for unit $i$.
Estimating $\tau$ by least squares is the same as estimating the following regression function by least squares,
\[ Y_i=\mu+\tau W_i+\gamma \overline{W}_{C_i}+\beta X_i+\delta \overline{X}_{C_i}+\eta_i,\]
 $\overline{W}_c$ is the cluster average of the treatment for cluster $c$, and similar for $\overline{X}_c$. This equivalence has been known since \citep*{mundlak1978pooling}.

\citep*{arkhangelsky2018role} build on the Mundlak insight, still in the clustered setting, by making the unconfoundedness assumption that 
\[ W_i\ \indep\ \Bigl(Y_i(0),Y_i(1)\Bigr)\ \Bigl|\ X_i,\overline{X}_{C_i},\overline{W}_{C_i}.\]
Implicitly this uses the two averages $\overline{X}_{C_i}$ and $\overline{W}_{C_i}$ as proxies for the differences between the clusters. This idea is related to \citep{altonji2005cross}, who also use exchangeability to control for unobserved heterogeneity. 
Given this uconfoundedness assumption, one can then adjust for differences in  $( X_i,\overline{X}_{C_i},\overline{W}_{C_i})$ in a flexible way, through non-parametric adjustment methods, possibly in combination with inverse propensity score weighting. 
\citep*{arkhangelsky2018role} then generalize this by assuming that 
\[ W_i\ \indep\ \Bigl(Y_i(0),Y_i(1)\Bigr)\ \Bigl|\ X_i,S_{C_i},\]
where the sufficient statistic $S_c$ captures the relevant features of the cluster, possibly including distributional features such as the average of $W_i$ in the cluster, but also other averages such as the average of the product of $X_i$ and $W_i$ in the cluster.

\citep*{arkhangelsky2022doubly} extend these ideas from the clustered cross-section case to the panel data case. They focus on the no-dynamics case where the potential outcomes are indexed only by the binary contemporaneous treatment.
In panel data settings, an alternative to two-way fixed effect regressions is the least squares regression 
\[ Y_{it}=\tau \ddot{W}_{it}+\varepsilon_{it},\]
where $\ddot{W}_{it}$ is the double difference
\[ \ddot{W}_{it}=W_{it}-\overline{W}_{i\cdot}-\overline{W}_{\cdot t}+\overline{\overline{W}},\]
with 
\[\overline{W}_{i\cdot}=\frac{1}{T}\sum_{t=1}^T W_{it},\quad
\overline{W}_{\cdot t}=\frac{1}{N}\sum_{i=1}^N W_{it},\quad \mathrm{and}\quad 
\overline{\overline{W}}=\frac{1}{NT}\sum_{i=1}^N\sum_{t=1}^T   W_{it}.
\]
See, for example, 
\citep{vogelsang2012heteroskedasticity}.
\citep{wooldridge2021two} shows the same estimator can be obtained by through what he calls the Mundlak regression
\[ Y_{it}=\tau {W}_{it}+\gamma \overline{W}_{i\cdot}+
\delta\overline{W}_{\cdot t}+ \varepsilon_{it}.\]

\citep*{arkhangelsky2022doubly} postulate the existence of a known function $S_i(W_{i1},\ldots,W_{iT})$ that captures all the relevant components of the assignment vector $\underline{W}_i=(W_{i1},\ldots,W_{iT})$ (and possibly other covariates, time-varying or time-invariant). Given this balancing statistic, they assume that the potential outcomes are independent of the treatment assignment vector given this balancing statistic:
\begin{equation}\label{eq:ai} \underline{W}_i\ \indep\  Y_{it}(w)\ \Bigl|\ S_i.\end{equation}
Consider the case where the balancing statistic is fraction treated periods, $S_i=\overline{W}_i$. The unconfoundedness assumption in (\ref{eq:ai})  implies that one can compare treated and control units in the same period, as long as they have the same fraction of treated periods over the entire sample. More generally $S_i$ could capture both the fraction of treated periods, as well as the number of transitions between treatment and control groups.

The estimator proposed by \citep*{arkhangelsky2022doubly} has a built-in robustness property: it remains consistent if the two-way model is correctly specified or the unconfoundedness given $S_i$ holds. As a result, it does not require researchers to commit to a single identification strategy. This approach creates a link between the TWFE literature and the design-based analysis we discuss in Section \ref{section:design}.

\subsection{Negative Controls, Proxies, and Deconvolution}\label{proxies}

The results in \citep{arkhangelsky2022doubly} show how to use panel data to construct a variable that eliminates the unobserved confounding. A related but different strategy is to use a panel to construct a set of proxy measures for the unobservables. If these proxy measures do not directly affect either outcomes or treatments, then this restriction can be used for identification. In biostatistics, such proxy variables are called negative control variables. To emphasize the connections between this literature and economic applications, we use these two terms, proxy variables and negative controls, interchangeably. 
In biostatistics a recent literature focuses on non-parametric identification results for average treatment effects that are based on negative controls (\citealp{sofer2016negative, shi2020multiply}). See \citep{ying2021proximal} for an introductory article. This literature is closely connected to econometric literature on non-parametric identification with measurement error (\citealp{hu2008instrumental}) and the CIC model (\citealp{athey2006identification}).
In a DID setting one can view the pre-treatment outcomes as proxies or negative controls in the sense of this literature.
Recently, these arguments have been extended to prove identification results for a class of panel data models in \citep{deaner2021proxy}. 

Proxy variables have a long history in economics. In early applications \citep{griliches1977estimating,chamberlain1977education} use data on several test scores to estimate returns to schooling accounting for unobserved ability (see also \citealp{deaner2021many}). Using modern terminology, these test scores serve as negative controls.  Versions of these strategies have also been successfully used in the traditional panel data literature. For example, \citep{holtz1988estimating} use data on past outcomes to estimate a dynamic linear panel data model with interactive fixed effects with a finite number of periods. They achieve this by eliminating the interactive fixed effects via a quasi-differencing scheme, which is called a bridge function in the negative control literature (see \citealp{imbens2021controlling,ying2021proximal}).

A similar idea is used in \citep{freyaldenhoven2019pre}, where the authors consider a setting with an unobserved confounder that can vary arbitrarily over $i$ and $t$. To eliminate this confounder, the authors assume the presence of a proxy variable that is affected by the same confounder but is not related to the treatment. As a result, one can eliminate the unobservables by subtracting a scaled proxy variable from the outcome of interest. The appropriate scaling is estimated using the pre-treatment data. In essence, this strategy is analogous to quasi-differencing and is another example of using bridge functions.

An important aspect of the negative control literature, which it shares with most of the methods discussed in this survey, is that it aims to isolate and eliminate the unobserved confounders rather than identify causal effects conditional on unobservables.  Alternatively, one can obtain identification under different distributional assumptions that connect the unobservables to outcomes and treatments using general deconvolution techniques. This approach has been successfully employed to answer causal questions in linear panel data models (\citealp{bonhomme2011recovering,arellano2011identifying}) and nonlinear quantile panel data models (\citealp{arellano2016nonlinear}), but so far has not been widely adopted by a broader causal community. 

\subsection{Combining Experimental and Observational Data}

Another direction this literature has explored is the combination of experimental and observational data.
\citep{athey2020combining} study the case with an experimental data set that has observations on short-term outcomes, and an observational sample that has information on the short-term outcome and the primary outcome. A key assumption is that the observational sample has an unobserved confounder that leads to biases in the comparison of the short-term outcome by treatment group. The experimental data allows one to remove the bias and isolate the unobserved confounder, which then can be used to eliminate biases in the primary outcome comparisons essentially as a proxy variable as discussed in the previous section.
See also \citep{ghassami2022combining, imbens2021controlling, kallus2020role}.

\section{Design-based Approaches to Estimation and Inference}\label{section:design}

An issue that features prominently in the recent panel data literature but is largely absent in the earlier one is a re-interpretation of the uncertainty in the estimates as coming from the stochastic nature of the causal variables. In most empirical analyses in economics and in most of the methodological literature in econometrics, uncertainty is assumed to be arising from sampling variation. This is a natural perspective if, say, we have data on individuals that can be at least approximately viewed as a random sample from a well-defined population. Had we sampled a different set of individuals, our estimates would have been different, and the standard errors reflect the variation that would have been seen if we repeatedly obtained different random samples from that population. This sampling-based perspective is still a natural one in panel data settings when the units can be viewed as a sample from a larger population, {\it e.g.,} individuals in the Panel Study of Income Dynamics or the National Longitudinal Survey of Youth.

The sampling-based perspective is less natural in cases where the sample is the same as the population of interest. This is quite common in panel data settings, for example, when we analyze state-level data from the United States, or country-level data from regions of the world, or all firms in a particular class. It is not clear why viewing such a sample as a random sample from a population is appropriate.
Researchers have struggled with interpreting the uncertainty of their estimates in that case.
Manski and Pepper write in their analysis of the impact of gun regulations with data from the fifty US states: ``measurement of
statistical precision requires specification of a sampling process that generates the data. Yet we are unsure what type of sampling process would be reasonable to assume in this
application. One would have to view the existing United States as the sampling realization of a random process defined on a superpopulation of alternative nations.'' (\citealp{manski2018right}, p. 234).

An alternative approach to formalizing uncertainty focuses on the random assignment of causes, taking the potential outcomes as fixed.
This approach has a long history in the analysis of randomized experiments 
({\it e.g.,} \citep{fisher1937design, neyman1923}), where the justification for viewing the causes as random is immediate. For modern discussions, see \citep{imbens2015causal, rosenbaum2023causal}.
Recently these ideas have been used to capture uncertainty in observational studies, see
\citep{abadie2020sampling, abadie2023should}. The justification in panel data settings is not always quite as clear.
Consider one of the canonical applications of synthetic control methods to estimate the causal effect of German re-unification in 1989 on West German Gross Domestic Product (GDP). 
A design-based approach would require the researcher to contemplate an alternative world where either other countries would have joined with East Germany or an alternative world where the reunification with West Germany would have happened in a different year. Both are difficult to contemplate. On the other hand, a sampling-based approach would require the researcher to consider a world with additional countries that could experience a unification event, which again is not an easy task. 

Design-based perspective has interesting connections with the econometric panel data literature. One such connection comes from the part of the panel data analysis that treats fixed effects as parameters rather than realizations of unobserved random variables, thus tying the inference to a particular set of observed units. The uncertainty in these models comes from errors, which we can interpret as realizations of {\it unobserved} treatments. In contrast, the design-based uncertainty comes from realizations of the observed treatment. A different connection is with panel data literature on random effects, which assumes that unit-level heterogeneity is uncorrelated with the regressors. This assumption is unlikely to hold in observational studies, but it holds by design in experiments. Again, the difference in analysis comes from fixing different quantities. The traditional analysis fixes the covariates and focuses on the distribution of the unit-level heterogeneity, whereas the design-based analysis does the opposite.

\subsection{The TWFE Estimator in the Staggered Adoption Case with Random Adoption Dates}

As an example of a design-based approach
\citep{athey2021design} analyze the properties of the TWFE estimator under assumptions on the assignment process in the staggered adoption setting, keeping the potential outcomes fixed. In that case the assignment process is fully determined by the distribution of the adoption date. 
\citep{athey2021design} derive the randomization-based distribution of the TWFE estimator under the random assignment assumption alone and present an interpretation for the estimand corresponding to that estimator.
They show that as long as the adoption date is randomly assigned, the estimand can be written as a linear combination 
of average causal effects on the outcome in period $t$ if assigned adoption date $a'$ relative to being assigned adoption date $a$:
\begin{equation}\label{eq:athey_imbens_2}
\tau^{a,a'}_t=\frac{1}{N}\sum_{i=1}^N\Bigl( Y_{it}(a')- Y_{it}(a)\Bigr),
    \end{equation}
with the weights summing to one but generally including negative weights.

\citep{athey2021design}  show the implications for the estimand of
the assumption that there is no anticipation of the treatment (so that the potential outcomes are invariant to the future date of adoption). 
They also show how the interpretation of the estimand changes further under the additional assumption that there are no dynamic effects so that the potential outcomes only depend on whether the adoption has taken place or not, but not on the actual adoption date. \citep{rambachan2020design} discuss the implications of variation in the assignment probabilities and the biases this can create.

\subsection{Switchback Designs}

One design that has recently received considerable attention after a long history is what 
\citep{cochran1939long} called the {\it rotation experiment}, and what more recently has been referred as a {\it switchback experiment} in \citep{bojinov2022design}
or {\it crossover experiment} in 
\citep{brown1980crossover}.
In such experiments units are assigned to treatment or control in each of a number of periods, with individual units potentially switching between treatment and control groups. Such experiments were originally used in agricultural settings, where, for example, cattle were assigned to different types of feed for some period of time. Using each unit as its own control can substantially improve the precision of estimators compared to assigning each unit to the treatment or control group for the entire study period. Such designs have become popular in tech company settings to deal with spillovers. For example, Lyft and Uber often randomize markets to treatment and control groups, with the assignment changing over time.

\subsection{Experimental Design with Staggered Adoption}

This subsection focuses on the design of experiments where the adoption date, rather than the treatment in each period,  is randomly assigned. 
Early studies, including
 \citep*{hemming2015stepped,
hussey2007design,
barker2016stepped} focused on simple designs, such as those where a constant fraction of units adopted the treatment in each period after the initial period. Sometimes these designs suggested analyses that allowed for spillovers so outcomes for one or two periods after the adoption would be discarded from the analyses if the focus was on the average treatment effect.

 \citep{xiong2019optimal} focused on the question of optimally choosing the fraction of adopters in each period and showed that instead of it being constant, it was initially small and then larger for some periods, after which it declined again.  \citep{bajari2023experimental} discuss randomization-based inference for some of these settings and present exact variances for some estimators.

 \subsection{Design with Dynamic Effects}

 \citep{bojinov2021panel} propose unbiased estimators and derive their properties under the randomization distribution. They allow for dynamics in the treatment effects and essentially unrestricted heterogeneity. They also discuss the biases of the conventional TWFE specifications in their setting.  \citep{bojinov2022design} discuss optimal design from a minimax perspective, allowing for carryover effects where the treatment status in recent periods may affect current outcomes.

\subsection{Robust Methods}

The design-based approach to estimation and inference is natural in the context of randomized experiments. However, in practice, applied researchers continue using conventional panel data methods, such as TWFE, even with experimental data (\textit{e.g.}, \citep{broda2014economic,colonnelli2022corruption}). There are multiple practical reasons for this: one can believe that the experiment's description is inconsistent with how it was implemented or think that the TWFE estimator is more precise. These concerns are even more salient in quasi-experimental environments where the data does not come from an experiment, but it is appealing to treat it as such (\textit{e.g.}, \citep{borusyak2022non}). 

To address these issues \citep{lihua2021} propose a version of the TWFE estimator that incorporates the design information. The key property of this method is that it delivers a consistent estimator even if the design assumptions do not hold as long as the TWFE model is correctly specified. Algorithmically, it amounts to estimating a weighted version of the standard TWFE model:
    \begin{equation}\label{estimator:wtwfe}
(\hat\tau^{rob},\hat\alpha,\hat\beta)=\arg\min_{\tau,\alpha,\beta}
\sum_{i=1}^N\sum_{t=1}^T \left( Y_{it}-\alpha_i-\beta_t-\tau W_{it}\right)^2\omega_i,
\end{equation}
where the weights $\{\omega_i\}_{i=1}^n$ are constructed using the information about the design.  In particular, in environments with staggered design, this amounts to estimating a duration model for the treatment adoption time. See \cite{shaikh2021randomization} for a related approach of using duration models for inference in staggered adoption designs. 

\section{Open Questions for Further Research}\label{section:future}

Here we discuss some open questions in the current causal panel data literature.

\subsection{Modeling Dynamics in Potential Outcomes}

The recent panel data literature has only paid limited attention to dynamic treatment effects, compared to the earlier literature (\citealp*{heckman2007dynamic, anderson1981estimation} (see \citep{abbring2007econometric} for an overview), as well as compared to its importance in practice.
For example, a curious feature of many of the current methods, including factor models and synthetic control methods, is that they pay essentially no attention to the time-ordering of the observations. If the time labels were switched, the estimated causal effects would not change. This seems implausible. Suppose one has data available for $T_0$ pre-treatment periods. For many of the methods, the researcher would be indifferent between having available the first $T_0/2$ pre-treatment period versus the second $T_0/2$ pre-treatment observations, whereas in practice, one would think that the more recent data would be more valuable.

It seems likely the current literature will take the dynamics more seriously. One direction may be to follow
\citep{robins1986new} and a number of follow-up studies that developed a sequential unconfoundedness approach. \cite{viviano2021dynamic} discuss this approach in economic contexts and propose an implementation that combines traditional linear models with modern balancing approaches. This analysis, however, ignores unobserved heterogeneity, which is central to the current empirical practice. See also \citep{brodersen2015inferring, ben2023estimating, masini2022counterfactual, masini2021counterfactual} for studies that take the time series dimension of these settings more seriously. Other recent work includes \citep{han2020identification, brown2023dynamic}.

\subsection{Validation}

\citep{lalonde1986evaluating} has become a very influential paper in the causal inference literature because it provided an experimental data set that could be used to validate new methods for estimating average causal effects under unconfoundedness. There are few longer panel data sets that can deliver the comparisons for validating the various new methods. However, there are methods that can be used to assess the performance of proposed estimators with purely observational data. 
An early paper with suggested tests is
 \citep{heckman1989choosing}. Currently, many approaches in panel data rely on placebo tests where the researcher pretends the treatment occurred some periods prior to when it actually did; the researcher then estimates the treatment effect for these periods where, in the absence of anticipation effects, the treatment effect is known to be zero. Finding estimates close to zero, both substantially and statistically, is then taken as evidence in favor of the proposed methods. See for examples \citep*{imbensrubinsacerdote} and \citep{abadie2014}. This strategy, however, relies on strict exogeneity and can backfire in models where the selection into treatment is based on shocks to past outcomes, as discussed in Section \ref{subsec:selection}. See \cite{arkhangelsky2023synth} for a particular illustration of this point and a discussion of alternative validation strategies.

\subsection{Connections with Macroeconomics}
\citep{nakamura2018identification} discuss the impact that ideas from the causal inference literature had on empirical research in macroeconomics. At the same time, the causal inference literature itself can benefit from incorporating macroeconomic ideas, which are particularly relevant in applications with panel data. For example, in Section \ref{notation}, we discuss that Lucas's critique can be relevant for the interpretation of causal quantities in applications in microeconomics. More broadly, panel data sets allow us to connect micro-level and aggregate time-series variation, providing identification strategies for aggregate effects (e.g., \citealp{gabaix2020granular}) as well as local-level effects (e.g., \citealp{arkhangelsky2019policy}). \cite{wolf2023missing} shows how to combine credible micro and macro evidence to analyze policy-relevant counterfactual in macroeconomic models.  We view this as an attractive area of future research.

\subsection{Bridging Unconfoundedness and the TWFE Approach}

Much of the discussion on unconfoundedness and the TWFE model has been framed in terms of a choice. It is difficult to imagine that a clear consensus will emerge, and finding practical methods that build on both approaches would be useful.

 \subsection{Continuous Treatments}

Much of the recent literature has emphasized the binary treatment case. This has led to valuable new insights, but it is clear that many applications go beyond the binary treatment case. 
There is a small literature studying these cases, including
\citep{callaway2021difference} and \citep{de2023two}, and earlier work in the cross-section setting, {\it e.g.,} \citep{imbens2000}, but more work is needed. Note that the earlier econometric panel data literature did not distinguish between settings where the variables of interest were binary or continuous.

\section{Recommendations for Empirical Practice}\label{recommendations}

The recent literature has greatly expanded the set of methods available to empirical researchers in social sciences in settings that are important in practice. This survey is an attempt to put these methods in context and show the close relationship between various approaches, including two-way-fixed-effect and synthetic control methods, to provide practitioners with additional guidance on when to use the various methods.

\subsection{The Blocked Assignment Setting}

Although the standard TWFE estimator (simplifying to the double difference or DID estimator in the special case with the blocked assignment) continues to be widely used, there are now methods available that have superior properties in settings with both cross-section and time dimensions at least modestly large. (In cases with few units and few time periods, there may not be enough information in the data to go beyond the simpler methods.) These methods relax the parallel trends assumption that is unattractive both from a conceptual perspective (because it is tied to a particular functional form) and from a practical perspective (because it is unlikely to hold over long periods of time). Some of the new methods allow for factor structures that generalize the two-way fixed effect setup. Others use synthetic control approaches, sometimes in combination with fixed effects. While none of these methods is likely to dominate uniformly, preliminary simulation evidence in the blocked assignment case ({\it e.g.,} \cite{arkhangelsky2021synthetic}) suggests that many of them dominate TWFE in realistic settings. Recent results in \cite{arkhangelsky2023synth} also suggest that some of these methods, in particular those based on synthetic control, dominate TWFE in settings with more complicated selection mechanisms. 

\subsection{The Staggered Adoption Case}

The staggered adoption case, common in empirical work, opens up new opportunities for estimation strategies (exploiting the variation in adoption times), but also forecloses some options (the standard synthetic control estimator).
Some of the recent proposals modify the TWFE estimator and relax the parallel trends assumptions by limiting the comparisons between treated and control outcomes to a subset of the set of possible comparisons. This subset may avoid comparisons distant in time, avoid the use of units that are to be treated at a future date as controls, or, in contrast, avoid the use of units that are never treated. In all cases there is an asymmetry in the way treated outcomes and control outcomes are used that does not appear to do justice to the {\it ex ante} arbitrariness in the treatment versus control labels. There have not been systematic simulation studies that are informative about realistic settings. Nevertheless, we expect that methods that model both treated and control potential outcomes, implying models for both control outcomes and treatment effects, taking account of dynamic effects as the earlier panel literature did more carefully, will be the most effective.

\bibliography{references}
\end{document}